\documentclass[a4paper,11pt]{article}
\pdfoutput=1

\usepackage{jheppub}
\usepackage[...]{youngtab}
\usepackage{xfrac}
\usepackage{appendix}
\usepackage{amsfonts}
\usepackage{amssymb}
\usepackage{physics}
\usepackage[usenames,dvipsnames]{xcolor}
\usepackage{tabularx,multirow,array,amsmath,mathalfa}
\usepackage{color}
\usepackage{ccaption}
\usepackage{mathtools}
\usepackage{bbold}
\usepackage{kantlipsum}
\usepackage{slashbox}
\usepackage{tikz}
\usepackage{pgfplots}
\pgfplotsset{compat=1.13}
\usepackage{braids}

\newtheorem{theorem*}{Theorem}

\allowdisplaybreaks

\usepackage{changepage}
\usepackage[framemethod=TikZ]{mdframed}
\usepackage{lipsum}
\usepackage{titlesec}
\mdfdefinestyle{Frame-1}{%
    linecolor=black,
    outerlinewidth=2pt,
    roundcorner=20pt,
    innerrightmargin=15pt,
    innerleftmargin=15pt,
    backgroundcolor=gray!30!white}	
\definecolor{Red}{rgb}{1,0,0}

\usepackage{graphicx}
\makeatletter

\newcommand*\bigcdot{\mathpalette\bigcdot@{.5}}
\newcommand*\bigcdot@[2]{\mathbin{\vcenter{\hbox{\scalebox{#2}{$\m@th#1\bullet$}}}}}
\makeatother
\usepackage{pgf}
\usepackage{tikz}
\usepackage[utf8]{inputenc}
\DeclarePairedDelimiter\floor{\lfloor}{\rfloor}
\usetikzlibrary{arrows,automata}
\usetikzlibrary{positioning}
\tikzset{state/.style={rectangle, rounded corners, draw=black, very thick, minimum height=2em, inner sep=2pt, text centered,},}

\title{\boldmath {Colored HOMFLY-PT for hybrid weaving knot $\hat{W}_{3}(m,n)$}}

\author[a]{Vivek Kumar Singh}
\author[a]{,~Rama Mishra}
\author[b]{,~and P. Ramadevi}

\affiliation[a]{\it Department of Mathematics, IISER, Pune, India }
\affiliation[b]{\it Department of Physics, Indian Institute of Technology Bombay, Mumbai 400076, India}

\emailAdd{vivek.singh@fuw.edu.pl}
\emailAdd{r.mishra@iiserpune.ac.in}
\emailAdd{ramadevi@phy.iitb.ac.in}

\abstract{Weaving knots $W(p, n)$ of type $(p, n)$ denote an infinite family of hyperbolic knots which have not been addressed by the knot theorists as yet. Unlike the well known $(p,n)$ torus knots,  we do not have a  closed-form expression for HOMFLY-PT and the colored HOMFLY-PT for $W(p,n)$.  In this paper, we confine to a hybrid generalization of $W(3,n)$ which we denote as $\hat{W}_3(m,n)$ and obtain closed form expression for HOMFLY-PT using the Reshitikhin and Turaev method involving $\mathcal R$-matrices. Further, we also   compute $[r]$-colored HOMFLY-PT for $W(3,n)$.  Surprisingly, we observe that trace of the product of two dimensional $\hat{\mathcal{R}}$-matrices can be written in terms of infinite family of Laurent polynomials $\mathcal{V}_{n,t}[q]$ whose absolute coefficients has interesting relation to the  Fibonacci numbers $\mathcal{F}_{n}$. We also computed reformulated invariants  and the BPS integers in the context of topological  strings.  From our analysis, we propose that certain refined BPS integers for weaving knot $W(3,n)$ can be explicitly derived from the coefficients of Chebyshev polynomials of first kind.}

\keywords{}

\arxivnumber{}

\begin{document} 

\maketitle
\section{Introduction}
Distinguishing knots and links up to ambient isotopy is the central problem in knot theory. The main technique that a knot theorist uses is to compute some knot invariants and see if one of them can be of help. Over the last 35 years tremendous progress has been made in the development of several new knot invariants, starting with the Jones polynomial and the HOMFLY-PT polynomial\cite{HOMFLY, PT, de1986jones}. Recently even more sophisticated invariants such as Heegard-Floer homology groups \cite{manolescu2014introduction} and Khovanov homology groups \cite{bar2002khovanov} have been added to the toolkit. In the 1980s William Thurston's seminal result \cite[Corollary 2.5]{thurston1982three} that most knot complements have the structure of a hyperbolic manifold, combined with Mostow's rigidity theorem \cite[Theorem 3.1]{thurston1982three} giving uniqueness of such structures, establishes a strong connection between hyperbolic geometry and knot theory, since knots are determined by their complements. Indeed, any geometric invariant of a knot complement, such as the hyperbolic volume, becomes a topological invariant of the knot. Thus, investigating if data derived from the new knot invariants is related to natural differential geometric invariants becomes another natural problem.  In this direction `volume conjecture' is one of the most challenging open problem. This conjecture has been tested for torus knots but for hyperbolic knots it has been verified only for a handful of knots.  Weaving knots $W(p,n)$ of type $(p,n)$ for a pair of co-prime integers $p$ and $n$ are doubly infinite family of alternating, hyperbolic knots and share the same projection with torus knots. They can be thought of a prototype of hyperbolic knots. Thus an extensive study of this family of knots will provide an insight to `volume conjecture.' One of us in an earlier work\cite{mishra2017jones} have attempted recursive method of relating the HOMFLY-PT of $W(3,n)$. In a parallel paper\cite{RRV}, the closed form of HOMFLY-PT for $W(3,n)$ with explicit proof is provided.  In this paper, we study the {\it hybrid family of weaving knots} denoted by $(\hat{W}_{3}(m,n))$ . Here, we use the approach of Reshitikhin and Turaev to evaluate the colored polynomials for knots and obtained the closed form expression for HOMFLY-PT polynomial for hybrid weaving knots. Further, we have computed the $[r]$-colored HOMFLY-PT polynomial for $W(3,n)$ which agrees when $[r]=[1]$ with the  results in \cite{RRV}. Further we study the reformulated invariants in the context of topological string dualities and validate Oogur-Vafa conjecture\cite{OV, GV1, LMV}. Interestingly, we  show that certain BPS integers of weaving knot $W(3,n)$ can be written in the  Chebyshev coefficients of first kind.  

The paper is organized as follows:

In section~\ref{s.int}, we will review Reshitikhin and Turaev (RT) method of constructing knot and link invariants which involves $\mathcal{R}$-matrices. This is followed by the subsection~\ref{s12.int} where we present the  $\hat{\mathcal{R}}$-matrices in a  block structure form for a three strands braid. In section~\ref{s3.int}, we used the properties of quantum $\hat{\mathcal R}_{2 \times 2}$ matrices, we succeeded in writing a closed form expression of HOMFLY-PT polynomial for $(\hat{W}_{3}(m,n))$.  As a consequence , we showed the  relation to the infinite set  of Laurent polynomials called $\mathcal{V}_{n,t}[q]$ whose absolute coefficients are  related to Fibonacci numbers . Section~\ref{s4.int} deals with $[r]$-colored HOMFLY-PT for weaving knots $W(3,n)$.  Particularly, we could express the trace of product of 2 dimensional matrices  as a Laurent polynomial. We  explicitly calculate colored polynomials for weave knot up to representation $[r]= [3]$. In section~\ref{s5.int}, we  verify that the reformulated invariants from these weave knot invariants indeed respect Ooguri-Vafa conjecture. The concluding section~\ref{s6.int} contains summary  and related  challenging open problems. There are two appendices with explicit data on colored HOMFLY-PT~\ref{app1} and reformulated invariants~\ref{app2} for $W(3,n)$.

\section{Knot invariants from quantum groups}
\label{s.int}
Recall Alexander theorem which states that any knot or link  can be viewed as closure of $m$-strand braid. 	Hence the knot invariants can be constructed from the braid group $\mathcal B_m$ representations. The representations of the generators $\sigma_i$'s of $\mathcal B_m$: $$\mathcal R_1, \mathcal R_2, \ldots \mathcal R_i, \ldots \mathcal R_{m-1}$$ are derivable from the well-known universal $\check{\mathcal{R}}$-matrix of $U_q(\mathfrak{sl}_N)$ defined as 
\begin{equation}
\check{\mathcal{R}} = q^{\sum\limits_{i,j}C_{ij}^{-1}H_i\otimes H_j}
\prod_{\textrm{positive root }\alpha} \exp_q[( 1-q^{-1}) E_\alpha\otimes F_\alpha]\,, \label {univR}
\end{equation}
where $q$ is complex number,  $( C_{ij})$ is the Cartan matrix  and and $\{H_i,E_i,F_i\}$ are generators of $U_q(sl_N)$.  Braid group generators ${\mathcal R}_i$'s, depicted in \ref{RM},  in terms of (\ref{univR}) is
\begin{equation}
\label{Rmat}
\mathcal{R}_i = 1_{V_1}\otimes1_{V_2}\otimes\ldots\otimes P \check{\mathcal{R}}_{i,i+1} \otimes\ldots\otimes1_{V_m} \ \in \text{End}(V_1\otimes\ldots,\otimes V_m)~,
\end{equation}
where  $P$ denotes the permutation operation: $P(x\otimes y) = y\otimes x$. Notice that the subscript $i, i+1$ on the  universal quantum $\check{{\mathcal{R}}}$ in the above equation implies $\check{{\mathcal{R}}}$ acts only on  the modules $V_i$ and $V_{i+1}$ of the $U_q(sl_N)$. The quantum $\mathcal{R}_i$ matrices discussed in \cite{KirResh}, \cite{Rosso:1993vn, lin2010hecke, Liu:2007kv} provides a  braid group $\mathcal B_m$ representation. That is.,
\begin{eqnarray}
\begin{array}{rcl}
\pi: \mathcal B_m &\rightarrow&  \text{End}(V_1\otimes\ldots,\otimes V_m) ~,\\
\pi(\sigma_i) &=& \mathcal{R}_i.
\end{array}
\end{eqnarray}
Graphically the braid group generator $\mathcal{R}_i$ as follows:
\begin{equation}\label{RM}
\begin{picture}(850,140)(-250,-90)

\put(-70,20){\line(0,-1){90}}
\put(-40,20){\line(0,-1){90}}

\put(0,20){\line(0,-1){30}}
\put(30,20){\line(0,-1){30}}
\put(0,-40){\line(0,-1){30}}
\put(30,-40){\line(0,-1){30}}
\put(70,20){\line(0,-1){90}}

\put(0,-10){\line(1,-1){30}}
\multiput(30,-10)(-17,-17){2}
{\line(-1,-1){13}}

\put(-105,-25){\mbox{$\mathcal{R}_i$ \ = \ }}
\put(45,-25){\mbox{$\ldots$}}
\put(-25,-25){\mbox{$\ldots$}}
\put(-75,25){\mbox{$V_1$}}
\put(-45,25){\mbox{$V_2$}}
\put(-5,25){\mbox{$V_i$}}
\put(20,25){\mbox{$V_{i+1}$}}
\put(60,25){\mbox{$V_m$}}
\end{picture}
\end{equation}
Algebriacally these generators in terms of (\ref{univR}). These operators $\mathcal{R}_i$ obeys the following relations :
\begin{eqnarray}
{\mathcal{R}}_i{{\mathcal{R}}}_j&=&{{\mathcal{R}}}_j{{\mathcal{R}}}_i~~ \text{for}~ |i- j|>1~,
\end{eqnarray}
\begin{eqnarray}\label{braidp}
{{\mathcal{R}}}_i{{\mathcal{R}}}_{i+1}{{\mathcal{R}}}_{i} &=&{ {\mathcal{R}}}_{i+1}{{\mathcal{R}}}_i{{\mathcal{R}}}_{i+1},~\text{for}~\ i=1,\ldots,m-2.
\end{eqnarray}
  Graphically, the  equation (b) is equivalent to the third Reidemeister move. According to Reshetikhin-Turaev approach \cite{RT1, RT2} the quantum group invariant, known as $[r]$-colored HOMFLY polynomial of the knot $\mathcal{K}$ denoted by $H_{[r]}^{\mathcal{K}}$ is defined as follows:
\begin{equation}
\label{HMF1}
H_{[
r]}^{\mathcal{K}} = 
{}_q\text{tr}_{V_1\otimes\dots\otimes V_m}\left( \, \pi(\alpha_{\mathcal{K}}) \, \right),
\end{equation}
where $_q\text{tr}$ is the quantum trace ( \cite{Klimyk}) defined as follows:
\begin{equation}
\label{qtr}
_q\text{tr}_V(z) = \text{tr}_V(zK_{2\rho})~~~\forall z \in \text{End}(V),
\end{equation}
where $\vec{\rho}$ is the Weyl vector that can expressed in terms of simple roots $\vec{\alpha}_i$ is $2 \vec{\rho}=\sum_i a_i \vec{\alpha}_i$ and the $K_{2 \rho}$ is defined as
$$K_{2\rho} = K_1^{a_1} \, K_2^{a_2} \ldots K_{N-1}^{a_{N-1}}$$
where  $K_p = q^{\vec{\alpha}_p . {\mathbf H}}$ having Cartan generators $H_1, H_2, \ldots
H_{N-1}$.

Note that the universal $\check R$ matrix is not diagonal and makes the computations of knot invariants very cumbersome. There is a modified RT-approach\cite{ModernRT1,Mironov:2011ym, Anokhina:2013wka} where the braiding generators  can be written in a block structure form. This methodology gives a better control  and simplify the computation of knot invariants.  We will present the details of this modified RT method in the following section.

\subsection{ $\hat{\mathcal{R}}$-matrices with Block structure}\label{s11.int}
The modified RT approach fixes the block structure form for $\hat{\mathcal R}_i$'s  from the  study of the irreducible representation in the tensor product  of symmetric representations $\underbrace{[r]\otimes [r]\otimes \ldots \otimes [r]}_m$: 
\begin{eqnarray}\label{irrdec}
 {[r]^{\bigotimes}}^{m} &=& \bigoplus_{\alpha,~\Xi_{\alpha} \vdash m |r|} ({\rm dim} {\mathcal{M}}^{1,2 \ldots m}_{\Xi_{\alpha}}) ~ \Xi_{\alpha}~,
\end{eqnarray}
where $\Xi_{\alpha}$ denote the irreducible representations labeled by index $\alpha$. The repetition in the irreducible representation called multiplicity( an irreducible representation occurs more than once) denoted by $\mathcal{M}_{\Xi_{\alpha}}^{1,2, \ldots m}$ that keep track of the subspace of the highest weight vectors\footnote{
Note that  the Young diagram $\Xi_{\alpha}$  represented as $[\xi_1^\alpha,\xi_1^\alpha,\ldots,\xi_l^\alpha]$ partitioned by $\{ \xi_1^\alpha \geq \xi_2^\alpha \geq\ldots,\xi_{l-1}^\alpha \geq\xi_l^\alpha\geq 0\}$ , then the highest weights $\vec{\omega}_{\Xi_{\alpha}}$ of the corresponding representation are $\omega^{\alpha}_i= \xi_i^{\alpha}-\xi_{i+1}^{\alpha} \ \forall \, i=1,\dots,l$, and vice versa ${\xi_i^{\alpha}} = \sum_{k=i}^l \, \omega^{\alpha}_k$.} sharing same highest weights corresponding to Young diagram $\Xi_{\alpha}\vdash m |r|$\footnote{$\Xi_{\alpha} \vdash m |r|$ means a sum over all Young diagrams $\Xi_{\alpha}$ of the size equal to $m |r|$. Here,$ |r|$ is total number of boxes in the Young diagram $[r]={\tiny \underbrace{\yng(3)\ldots \yng(2)}_{r}}$}, which we indicate as $\Xi_{\alpha,\mu}$ with the index $\mu$, takes values $1,2,\ldots {\rm dim} \mathcal{M}_{\Xi_{\alpha}}^{1,2,\ldots m}$, keep track of the different highest weight vectors sharing the same highest weight $\vec {\omega}_{\Xi_{\alpha}}$. 

To evaluate quantum trace(\ref{qtr}), we need to write the states in weight space incorporating the multiplicity as well. There are several paths leading  to the state corresponding to the irreducible representations $\Xi_{\alpha}$. Pictorially depicted one such state in the weight space (see in (\ref{pic2})).
\begin{equation}
\hskip6cm
\begin{picture}(400,80)(-40,-32)\label{pic2}
\put(0,0){\line(-1,1){30}}
\put(-24,24){\vector(1,-1){0}}
\put(0,0){\line(1,1){30}}
\put(24,24){\vector(-1,-1){1}}
\put(-10,10){\line(1,1){20}}
\put(04,24){\vector(-1,-1){1}}
\put(-20,20){\line(1,1){10}}
\put(-10,30){\vector(-1,-1){6}}
\put(-37,35){\mbox{$[r]$}}
\put(-13,35){\mbox{$[r]$}}
\put(7,35){\mbox{$[r]$}}
\put(27,35){\mbox{$[r]$}}
\put(-27,9){\mbox{$\Lambda_{\alpha}$}}
\put(-22,-1){\mbox{$\Xi_{\alpha_1}$}}
\multiput(2,-2)(3,-3){3}{\circle*{2}}
\put(-9,-15){\mbox{$\Xi_{\alpha_2}$}}
\put(10,-10){\line(1,1){40}}
\put(44,24){\vector(-1,-1){0}}
\put(47,35){\mbox{$[r]$}}
\put(10,-10){\line(1,-1){10}}
\put(10,-10){\vector(1,-1){8}}
\put(17,-30){\mbox{$\Xi_{\alpha}$}}
\end{picture}
\end{equation}
 and algebraically it can written as
 \begin{equation}
\label{decomp}
\vert \left(\ldots \left(( [r] \otimes [r])_{\Lambda_{\alpha}} \otimes [r]\right)_{{\Xi}_{\alpha_1}} \ldots [r]\right)_{\Xi_{\alpha}}\rangle^{(\mu)} \equiv | \Xi_{\alpha}; \Xi_{\alpha,\mu}, \Lambda_{\alpha}\rangle
\cong~| \Xi_{\alpha,\mu},\Lambda_{\alpha}
\rangle \otimes |\Xi_{\alpha}\rangle~,
\end{equation}
where $[r]\otimes [r]=\oplus_{\alpha=0}^r \Lambda_{\alpha}\equiv  [2 r-\alpha,\alpha]$. For clarity, in this paper, we denote the $\hat{\mathcal{R}}_{i}$'s- matrices corresponding to $\Xi_{\alpha}$ as $\hat{\mathcal{R}}_i^{\Xi_{\alpha}}$.  Incidentally, the choice of state (\ref{decomp}) is an eigenstate of quantum
$\hat{\mathcal{ R}}^{\Xi_{\alpha}}_1$ matrix:
\begin{eqnarray}
\hat{\mathcal{R}}^{\Xi_{\alpha}}_1 \vert \Xi_{\alpha}; \Xi_{\alpha,\mu}, \Lambda_{\alpha} \rangle =
\lambda_{\Lambda_{\alpha},\mu}([r],[r])\vert \Xi_{\alpha}; \Xi_{\alpha,\mu}, \Lambda_{\alpha} \rangle \rangle\nonumber.
\end{eqnarray}
Hence we will denote the $\hat{\mathcal R}^{\Xi_{\alpha}}_1$ matrix which is diagonal in the above basis and the elements denoted by $\lambda_{\Lambda_{\alpha},\mu}([r],[r])$.
These elements are the braiding eigenvalues
whose  explicit form is \cite{Klimyk,GZ}
 \begin{equation}
 \label{evR1}
       \lambda_{\Lambda_{\alpha},\mu} ([r],[r])=
            \epsilon_{\Lambda_{\alpha},\mu} q^{\varkappa(\Lambda_{\alpha})-
4\varkappa([r])-r N}~,
        \end{equation}
where $\varkappa(\Lambda_{\alpha})=\tfrac{1}{2} \sum_j \alpha_j (\alpha_j+1-2j)$\footnote{The representation $\Lambda_{\alpha}$ whose Young diagram is denoted by $\alpha_1\geq \alpha_2\ldots,\geq\alpha_{N-1}$} is cut-and-join-operator eigenvalue of Young tableaux representation $\Lambda_{\alpha}$ that does not depend on the braid representation of the knot $\mathcal{K}$\cite{MMN,MMN1} and $\epsilon_{\Lambda_{\alpha},\mu}$ will be $\pm 1$.\footnote { The multiplicity subspace state $\Xi_{\alpha,\mu}$  is connected by $\Lambda_{\alpha}$ and zero otherwise.} From the eqn.(\ref{decomp},\ref{HMF1}), incorporating all the facts of  $\hat{\mathcal{R}}$-matrix and the decomposition of states (\ref{decomp})\footnote{In facts, the action of $\hat{\mathcal{R}}$-matrix acts  an identity operator on $|\Xi_{\alpha} \rangle$ and non-trivially on the subspace $\mathcal{M}_{\Xi_{\alpha}}^{1,2,\ldots m}$ and similarly on other way, the element $K_{2\rho}$ acts diagonally on $|\Xi_{\alpha}\rangle$ but as identity operator on subspace $\mathcal{M}_{\Xi_{\alpha}}^{1,2 \ldots m}$ as this space represent all possible highest weight vectors $\Xi_{\alpha,\mu}$ with the same weight $\vec {\omega}_{\Xi_{\alpha}}$.}, the unreduced [r]-colored HOMFLY-PT  will become 
\begin{eqnarray}
H_{[r]}^{\star \mathcal{K}} \left(q,\,A=q^N \right)
&=&
\text{tr}_{V_1\otimes\dots\otimes V_m}\left( \, \pi(\alpha_{\mathcal{K}}) \, K_{2\rho} \, \right) \ = \  \sum_{\alpha} \text{tr}_{\mathcal{M}_{\Xi_{\alpha}}^{1,2 \ldots m}}\left( \, \pi(\alpha_{\mathcal{K}}) \, \right) \cdot \text{tr}_{\Xi_{\alpha}}\left( \, K_{2\rho} \, \right) \nonumber\\
&=&  \sum_{\alpha} \text{tr}_{\mathcal{M}_{\Xi_{\alpha}}^{1,2 \ldots m}}\left( \, \pi(\alpha_{\mathcal{K}}) \, \right) \cdot S^*_{\Xi_{\alpha}}~=\sum\limits_{\alpha, \Xi_{\alpha}\vdash m |r|} S^{*}_{\Xi_{\alpha}} C_{\Xi_{\alpha}}^{\mathcal{K}},
\label{HOMFLY}
\end{eqnarray}
where $\Xi_{\alpha}$ represent the  irreducible representations in the product $[r]^{\otimes m}$, $m$ stands for number of braid strands, $[r]$ denotes the representation on each strand, $C_{\Xi_{\alpha}}$ having the trace of product of all $\hat{\mathcal{R}}$-matrices, and $S^*_{\Xi_{\alpha}}$ is the quantum dimension of the representation $\Xi_{\alpha}$ whose explicit form is given in terms of Schur polynomials \cite{Mironov:2011aa, Dhara:2018wqe}. Note that the notation $H_{[r]}^{\star \mathcal{K}}$ denote unreduced  HOMFLY-PT of knot $\mathcal{K}$. The reduced [r]- colored HOMFLY-PT($H_{[r]}^{\mathcal{K}}$) is obtained by dividing the $[r]$-colored unknot invariant ($H_{[r]}^{unknot}$) i.e 
\begin{eqnarray}\label{NHOM}
H_{[r]}^{\mathcal{K}}&=&\frac{H_{[r]}^{\star \mathcal{K}}}{H_{[r]}^{unknot}}=\frac{H_{[r]}^{\star \mathcal{K}}}{S^{*}_{[r]}}.
\end{eqnarray}
 For clarity, we present the invariants of knots obtained from the simplest two-strand braids using this method in the following subsection.
\subsubsection{$[r]$-colored HOMFLY-PT  polynomial for closure of two strand braids}
\begin{figure}[h]
\centering
\includegraphics[scale=.6]{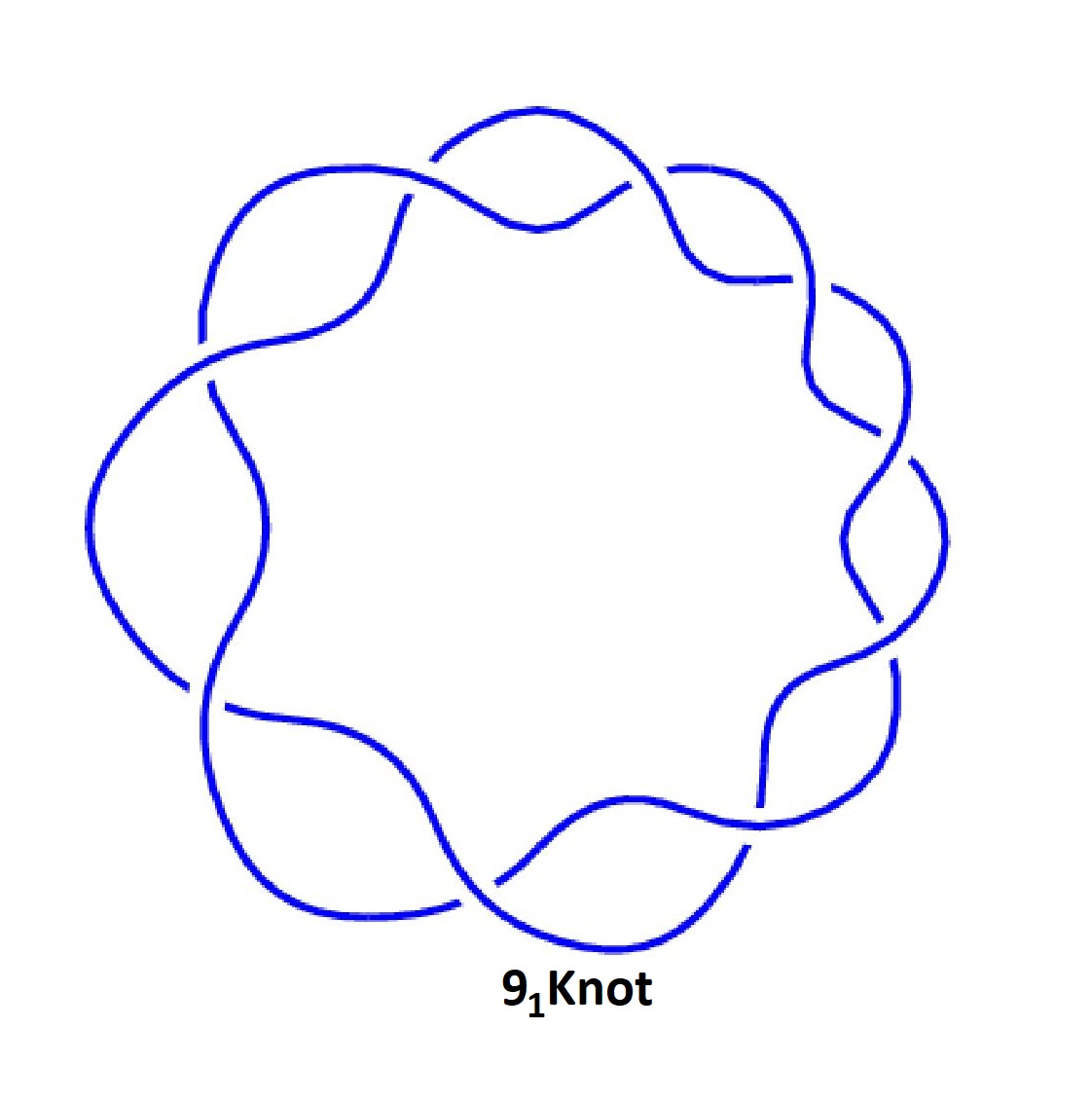}
\caption{ Torus knot $T_{(2,9)}={\bf 9_{1}}$ knot}
\label{TK}
\end{figure}

We will illustrate  the $[r]$-colored HOMFLY-PT for  knot $\mathcal{K}$ carrying symmetric representation $[r]$ obtained from the braid word $\sigma_1^{n}$ (\ref{2st}) where $n$ is odd integer. These knots known as torus knot $T_{(2,n)}$ and we have drawn as a example in Fig.\ref{TK}.The irreducible representation $\Lambda_{\alpha}$ in the tensor product of $[r]\bigotimes[ r]=\bigoplus_{\beta=0}^{r} \Lambda_{\alpha}=\bigoplus_{\beta=0}^{r} [2 r-\alpha,\alpha]$ has no multiplicity. Note that each irreducible representation occurs only once. So the $\hat{\mathcal R}$ are only eigenvalues and not matrices.

\begin{equation}
\hskip0cm
\begin{tikzpicture}{\label{2st}}
\braid[number of strands=2,rotate=90] (braid) a_1 a_1 a_1 a_1 a_1;
\end{tikzpicture}
\ldots n
\end{equation}

Hence, using eqns.(\ref{HOMFLY}, \ref{evR1}), we can obtain colored HOMFLY-PT $H_{[r]}^{\mathcal{K}}(q,A=q^N)$ which involves the single diagonal $\hat{\mathcal{R}}$-matrix i.e $\hat{\mathcal{R}}_{1}^{n}$ whose explicit entries depicted from eqn.(\ref{evR1}). 
\begin{eqnarray}{\label{egn}}
\lambda_{\Lambda_{\alpha}}([r],[r]) &=& (-1)^{\alpha}q^{(2 r^2- r (2 \alpha +1)+\alpha(\alpha-1)-rN)}.
\end{eqnarray}
 The HOMFLY-PT for torus knot $T_{(2,n)}$ 
\begin{eqnarray}\label{HOMR}
H_{[r]}^{\mathcal{K}}(q,A)&=&\frac{\sum_{\alpha}\text{tr}_{\Lambda_{\alpha}}\hat{\mathcal{R}}^{n}S^*_{\Lambda_{\alpha}}}{ S^*_{[r]}}=\frac{1}{ S^*_{[r]}}\sum_{\alpha} \lambda_{\Lambda_{\alpha}}([r],[r])^{n} ~ S^*_{ \Lambda_{\alpha}},\nonumber\\
&=&\frac{A^{(-r n))}}{ S^*_{[r]}}\sum_{\alpha=0}^{r} (-1)^{n\alpha} q^{(2 r^2- r (2 \alpha +1)+\alpha(\alpha-1)) n}
~S^*_{ \Lambda_{\alpha}}.
\end{eqnarray}
 The explicit form of quantum dimension $S^*_{ \Lambda_{\alpha}}$
 \begin{equation*}
S^*_{\Lambda_{\alpha}} = \frac{[N+\alpha-2]_q!\,[N+2r-\alpha-1]_q!\,[2r-2\alpha+1]_q}{[\alpha]_q!\,[2r-\alpha+1]_q!\,[N-1]_q!\,[N-2]_q!} ,
\end{equation*}
where the factorial is defined as $[n]_{q}! = \prod_{i=1}^n [i]_q$ with $[0]_{q}!=1$ and the $q$-numbers for our computation will be given as,
\begin{equation}
[n]_{q} = \frac{q^{n} - q^{-n}}{q^{1} - q^{-1}}.
\label{q-number}
\end{equation}
The explicit polynomial form for colors $[r]=[1]$ and $[r]=[2]$ for this knot $T(2,9)={\bf 9_1}$ are
\begin{eqnarray*}
H^{{\bf 9_1}}_{[1]}(q,A)&=&A^8+\frac{A^8}{q^8}-\frac{A^{10}}{q^6}+\frac{A^8}{q^4}-\frac{A^{10}}{q^2}-A^{10} q^2+A^8 q^4-A^{10} q^6+A^8 q^8,\\
H^{{\bf 9_1}}_{[2]}(q,A)&=&(A^{16}-A^{18}+A^{20}+\frac{A^{16}}{q^{16}}-\frac{A^{18}}{q^{12}}+\frac{A^{16}}{q^{10}}-\frac{A^{18}}{q^{10}}+\frac{A^{16}}{q^8}-\frac{A^{18}}{q^6}+\frac{A^{20}}{q^6}+\frac{A^{16}}{q^4}\\&&-\frac{2
A^{18}}{q^4}+\frac{A^{16}}{q^2}-\frac{A^{18}}{q^2}+A^{16} q^2-2 A^{18} q^2+A^{20} q^2+A^{16} q^4-2 A^{18} q^4+A^{16} q^6\\&&-2 A^{18} q^6+A^{20} q^6+2
A^{16} q^8-2 A^{18} q^8+A^{20} q^8+A^{16} q^{10}-2 A^{18} q^{10}+A^{20} q^{10}\\&&+A^{16} q^{12}-3 A^{18} q^{12}+A^{20} q^{12}+2 A^{16} q^{14}-3 A^{18}
q^{14}+A^{20} q^{14}+2 A^{16} q^{16}\\&&-2 A^{18} q^{16}+A^{20} q^{16}+A^{16} q^{18}-3 A^{18} q^{18}+2 A^{20} q^{18}+2 A^{16} q^{20}-3 A^{18} q^{20}+\\&&A^{20}
q^{20}+A^{16} q^{22}-2 A^{18} q^{22}+A^{20} q^{22}+A^{16} q^{24}-2 A^{18} q^{24}+A^{20} q^{24}+A^{16} q^{26}\\&&-2 A^{18} q^{26}+A^{20} q^{26}+A^{16}
q^{28}-A^{18} q^{28}-A^{18} q^{30}+A^{20} q^{30}+A^{16} q^{32}-A^{18} q^{32}).
\end{eqnarray*}
 
 If we go beyond two-strand braids, we need to deal with quantum $\hat{\mathcal R}_i's$ which could be matrices depending on the multiplicity sub-spaces. As our focus is on weaving knots $W(3,n)$ and their hybrid generalization, we will elaborate the steps of the modified RT method for three strands braid in the following section. Notice that, the braiding property eqn.(\ref{braidp}) means that both $\hat{\mathcal R}_1$ and $\hat{\mathcal R}_2$ cannot be simultaneously diagonal but related by a unitary matrix which can be identified with the $U_q(sl_N)$ Racah matrices.

\subsection{ $\hat{\mathcal{R}}$- matrices with Block structure for three strand braids}\label{s12.int}
 For three strand braids and each strands carrying the symmetric representation, 
the tensor product of representations $\underbrace{[r]\otimes [r]\otimes \ldots \otimes [r]}_m$ into the direct sum of irreducible representations($\Xi_{\alpha}$) is shown: 
\begin{eqnarray*}
\label{irrdec}
\bigotimes^3 [1] &=& [3,0,0]\bigoplus [1,1,1]\bigoplus 2 [2,1,0],\\
\bigotimes^3 [2]& =& [6,0,0]\bigoplus [3,3,0]\bigoplus  [4,1,1]\bigoplus 2 [5,1,0]\bigoplus 2 [3,2,1]\bigoplus 3 [4,2,0],\\
\bigotimes^3 [3]& =& [9,0,0]\bigoplus [7,1,1]\bigoplus  [5,2,2]\bigoplus  [4,4,1]\bigoplus  [3,3,3]\bigoplus 3 [8,1,0]\bigoplus  \\&&
2[4,3,2]\bigoplus 2 [6,2,1]\bigoplus 2 [5,4,0]\bigoplus 3 [7,2,0]\bigoplus 2 [5,3,1]\bigoplus 4 [6,3,0],\\
\ldots\\
\bigotimes^3 [r] &=&\sum_{\alpha}({\rm dim} {\mathcal{M}}^{1,2, 3}_{\Xi_{\alpha}}) \Xi_{\alpha},
\end{eqnarray*}
where  $\Xi_{\alpha}\equiv[{\xi_1}^{\alpha},{\xi_2}^{\alpha},{\xi_3}^{\alpha}]$ is such that ${\xi_1}^{\alpha}+{\xi_2}^{\alpha}+{\xi_3}^{\alpha}=3 r$ and ${\xi_1}^{\alpha}\geq {\xi_2}^{\alpha}\geq{\xi_3}^{\alpha} \geq 0 $.
Let us discuss the  path and block structure of $\hat{\mathcal{R}}$-matrix for irreducible representation [4,2,0]. Note that the multiplicity of the representation $[4,2,0]$ is equal to three which means there are three possible paths:
\begin{equation}
\begin{array}{llclclcl}
(i) & [2] & \rightarrow & [2] & \rightarrow & [4] & \rightarrow & [4,2,0] ~,\\
(ii) & [2] & \rightarrow & [2] & \rightarrow & [3,1] & \rightarrow & [4,2,0]~, \\
(iii) & [2] & \rightarrow & [2] & \rightarrow & [2,2] & \rightarrow & [4,2,0]. \\
\end{array}
\end{equation}
 Let us choose $\hat{\mathcal{R}}^{[4,2,0]}_1$ to be diagonal  whose entries defined by (\ref{evR1})
 \begin{equation}
 \label{R1}
       \lambda_{[2,2]} ([2],[2])= A^{-2},~        \lambda_{[3,1]} ([2],[2])= -A^{-2} q^2,~\lambda_{[4]} ([2],[2])=A^{-2} q^6.
\end{equation}
The explicit form of $\hat{\mathcal{R}}^{[4,2,0]}_1$
\begin{equation}
\hat{\mathcal{R}}^{[4,2,0]}_1=A^{-2} \left(\begin{array}{ccc} 1& 0&0 \\ \\  0 & -q^2&0\\ \\  0 & 0&q^6\end{array}\right).
\end{equation}
$\hat{\mathcal{R}}_2$ is defined as 
\begin{equation*}
\hat{\mathcal{R}}_2^{\Xi_{\alpha}}=\mathcal{U}^{\Xi_{\alpha}}\hat{\mathcal{R}}_1^{\Xi_{\alpha}}(\mathcal{U}^{ \Xi_{\alpha})^{\dagger}}.
\end{equation*}
Note that $\mathcal{U}^{\Xi_{\alpha}\dagger}$ denotes the conjugate-transpose of $\mathcal{U}^{\Xi_{\alpha}}$. This unitary matrix relate  two equivalent basis states for irreducible representation $[4,2,0]$  as shown below in: 
\begin{equation}\label{path}
\begin{picture}(200,75)(0,-35)
\put(0,0){\line(-1,1){30}}
\put(0,0){\line(1,1){30}}
\put(-15,15){\line(1,1){15}}
\put(-33,32){\mbox{$[2]$}}
\put(-3,32){\mbox{$[2]$}}
\put(27,32){\mbox{$[2]$}}
\put(-30,7){\mbox{$\Lambda_{\alpha}$}}
\put(-0,0){\line(1,-1){15}}
\put(0,-20){\mbox{$\Xi_{\alpha}=\{3[4,2,0] \}$}}
\put(58,0){\mbox{$\ \ \  \overset{\mathcal{U}^{[4,2,0]}}{\longrightarrow}$}}
\put(150,0){
\put(0,0){\line(-1,1){30}}
\put(0,0){\line(1,1){30}}
\put(15,15){\line(-1,1){15}}
\put(-33,32){\mbox{$[2]$}}
\put(-3,32){\mbox{$[2]$}}
\put(27,32){\mbox{$[2]$}}
\put(15,5){\mbox{$\Lambda_{\alpha'}$}}
\put(0,0){\line(1,-1){15}}
\put(8,-23){\mbox{$\Xi_{\alpha}=\{3[4,2,0] \}$}}
}
\end{picture}
\end{equation}
where, $\Lambda_\alpha ~\&~ \Lambda_\alpha' \in\{[4],[3,1],[2,2]\}$ and algebraically the transformation state for $\Xi_{\alpha}$ are: 
\begin{equation*}
\label{asis}
|\left(\left( [r]\otimes [r]\right)_{\Lambda_{\alpha}}\otimes [r]\right)_{\Xi_{\alpha}} \rangle \xrightarrow[\text{}]{\mathcal{U}^{\Xi_{\alpha}}} \vert \left([r] \otimes \left( [r]\otimes [r] \right)_{\Lambda_{\alpha'}} \right)_{\Xi_{\alpha}}\rangle~,
\end{equation*}
where the elements of the transformation matrix $\mathcal{U}^{\Xi_{\alpha}}$ related to quantum Racah coefficients discuss in details \cite{Itoyama:2012re, Dhara:2017ukv, Dhara:2018wqe}. For completeness, Racah matrix involving $\Xi_{\alpha}\equiv [{\xi_1}^{\alpha},{\xi_2}^{\alpha},{\xi_3}^{\alpha}]$ (whose Young diagram has three rows) can be identified as  $U_q(sl_2)$ Racah matrix:
\begin{eqnarray}
\label{symrac}
\mathcal{U}^{\Xi_{\alpha}\equiv [{\xi_1}^{\alpha},{\xi_2}^{\alpha},{\xi_3}^{\alpha}]}
&=&
U_{U_q(sl_2)}
\begin{bmatrix}
(r-{\xi_3}^{\alpha})/2 & (r-{\xi_3}^{\alpha})/2 \\
~&~\\
(r-{\xi_3}^{\alpha})/2 & ({\xi_1}^{\alpha}-{\xi_2}^{\alpha})/2.
\end{bmatrix}
\end{eqnarray}
The closed form expression of $U_q(sl_2)$ Racah coefficients \cite{KirResh}:
\begin{eqnarray*}
U^{U_q(sl_2)}_{j,l}
\begin{bmatrix}
j_1 &j_2\\
~&~\\
(j_3 & j_4 \end{bmatrix}&=&\sqrt{[2 j+1]_q [2 l+1]_q} (-1)^{j_1+j_2+j_3+j_4+j+l+1} \Delta(j_1, j_2, j)
\Delta(j_3,~j_4,j)\\&& \Delta(~j_4,~j_1, l) \Delta(~j_2,~j_3,l)F[j_1,j_2,j_3,j_4]~,\\
\end{eqnarray*}
where,
\begin{eqnarray*}
F[j_1,j_2,j_3,j_4]&=&\sum _{m\geq0}(-1)^m {[m+1]_q}!\{{[m-(j+~j_1+~j_2)]_q}! {[m-(j+~j_3+~j_4)]_q}! {[(m-(~j_1+~j_4+l))]_q}!\\&&{[m-(~j_2+~j_3+l)]_q}! {[(j+~j_1+~j_3+l)-m]_q}! ~{[(j+~j_2+~j_4+l)-m]_q}!\\&&~{[(~j_1+~j_2+~j_3+~j_4)-m]_q}!\}^{-1}\\
\Delta(a, b, c)&=&\sqrt{\frac{{[a-b+c]_q}!{[b-a+c]_q}!{[a+b-c]_q}!}{ {[a+b+c+1]_q}!}}~.
\end{eqnarray*}
Hence, from eqn.(\ref{symrac}) the explicit form of unitary matrix $\mathcal{U}^{[4,2,0]}$ defined as
\begin{equation}
\mathcal{U}^{[4,2,0]}=\left(
\begin{array}{ccc}
 -\frac{1}{1+\frac{1}{q^2}+q^2} & -\frac{q}{\sqrt{1+q^2+q^4}} & -\frac{\sqrt{1+q^2+q^4+q^6+q^8}}{1+q^2+q^4} \\
 -\frac{q}{\sqrt{1+q^2+q^4}} & -1+\frac{q^2}{1+q^4} & \frac{q^4 \sqrt{\left(1+\frac{1}{q^2}+q^2\right) \left(1+\frac{1}{q^4}+\frac{1}{q^2}+q^2+q^4\right)}}{\left(1+q^4\right)
\left(1+q^2+q^4\right)} \\
 -\frac{\sqrt{1+q^2+q^4+q^6+q^8}}{1+q^2+q^4} & \frac{q^4 \sqrt{\left(1+\frac{1}{q^2}+q^2\right) \left(1+\frac{1}{q^4}+\frac{1}{q^2}+q^2+q^4\right)}}{\left(1+q^4\right)
\left(1+q^2+q^4\right)} & -\frac{q^4}{1+q^2+2 q^4+q^6+q^8} \\
\end{array}
\right).
\end{equation}
Hence, the $\hat{\mathcal{R}}_{2}$ matrix for $[4,2,0]$ is
$$\hat{\mathcal{R}}^{[4,2,0]}_2=\mathcal{U}^{[4,2,0]} \hat{\mathcal{R}}^{[4,2,0]}_1\mathcal{U}^{[4,2,0]\dagger}.$$
Explicit form of quantum $\hat{\mathcal{R}}_i$'s and $\mathcal{U}$ can be similarly worked out for other irreducible representations to compute $[r]$-colored HOMFLY-PT for hybrid weaving knots.

\section{Hybrid weaving knot $\hat{W}_3(m, n)$}\label{s3.int}
\begin{figure}[h]
\centering
\includegraphics[scale=.6]{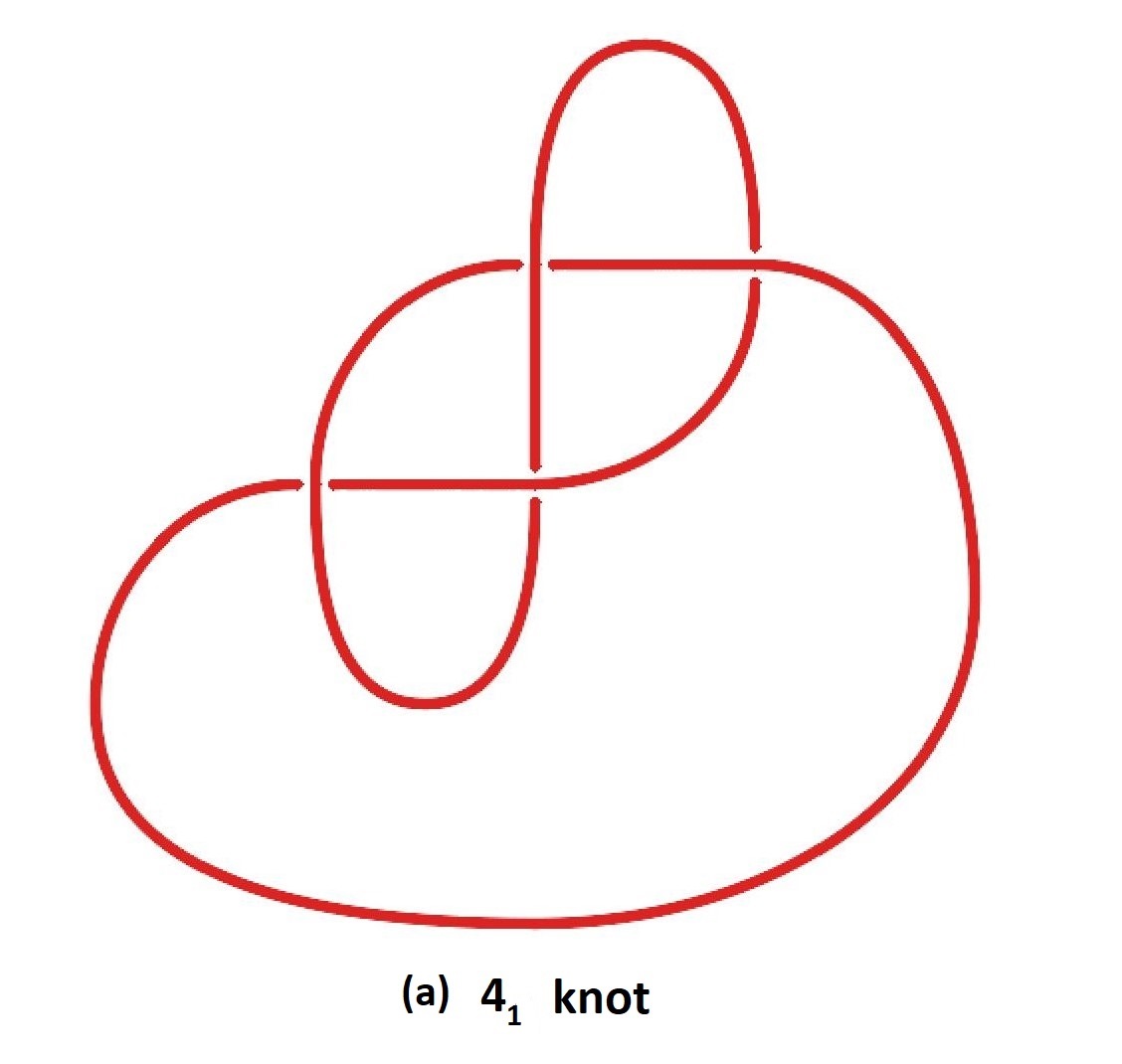}
\includegraphics[scale=.6]{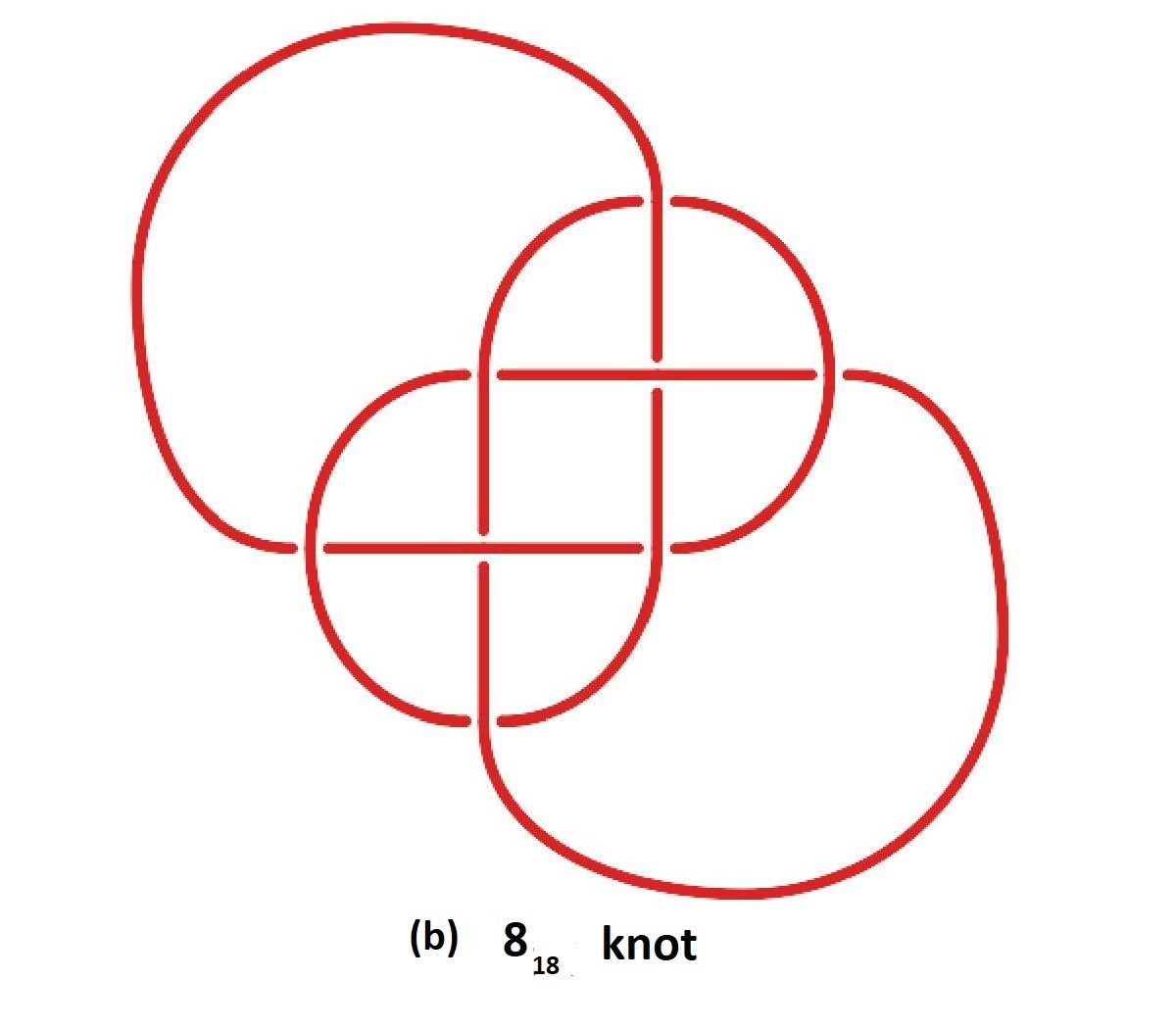}
\includegraphics[scale=.6]{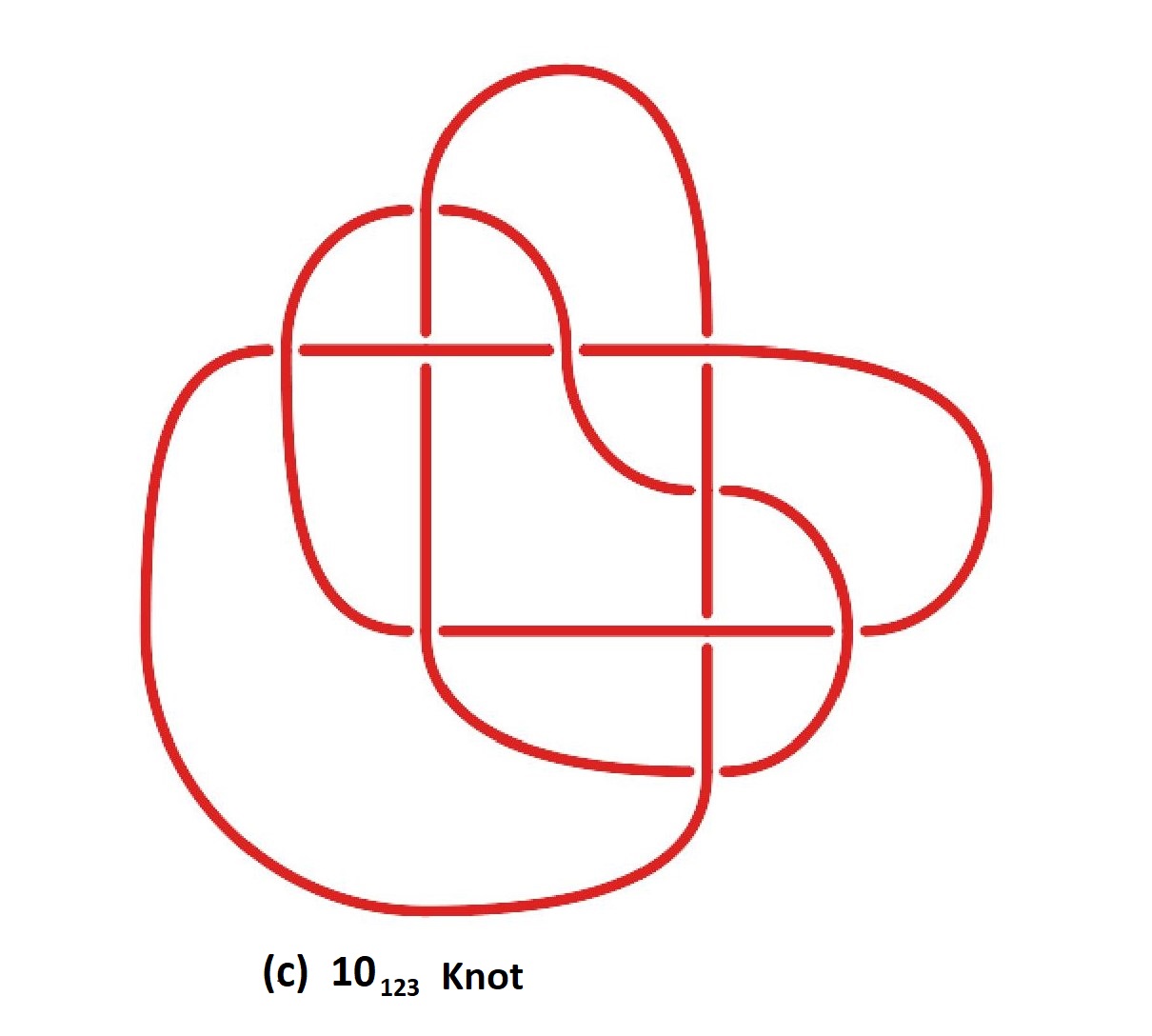}
\includegraphics[scale=.6]{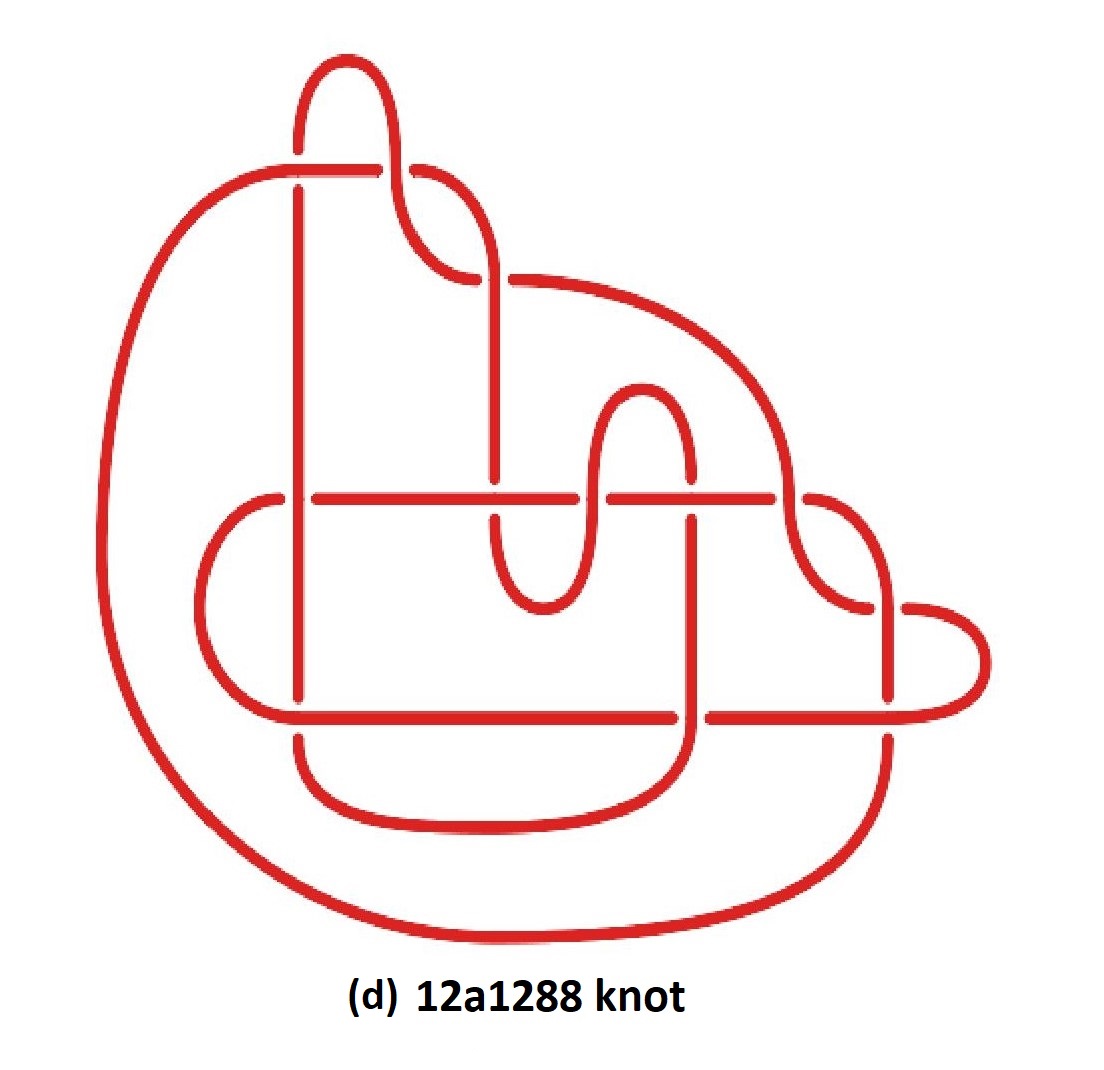}
\caption{Snappy diagram representation for hybrid knots\cite{katlas, indiana}: (a) $\hat{W}_3(2,1)={\bf 4_{1}}$, (b) $\hat{W}_3(4,1)={\bf 8_{18}}$ knot, (c) $\hat{W}_3(5,1)={\bf 10_{123}}$ knot, and (d) $\hat{W}_3(3,2)={\bf 12a1288}$ knot}
\label{HKD}
\end{figure}

In this section, we discuss the hybrid weaving knot obtained from closure of three-strand braid whose braid word is
$$(\sigma_1^{m} \sigma_2^{-m})^n$$ which is pictorially seen in (\ref{GWK}). Note that  the subscript 3 in $\hat{W}_3(m, n)$ indicates three-strand braid.
\begin{equation}
\begin{picture}(350,100)(-20,-50)\label{GWK} 

\put(40,30){\vector(1,0){35}}
\put(40,0){\vector(1,0){55}}
\put(40,-30){\vector(1,0){35}}
\put(75,-30){\vector(1,0){80}}
%
%
\put(75,30){\vector(1,0){40}}
\put(95,0){\vector(1,0){20}}
\qbezier(115,30)(120,30)(120,25)
\qbezier(115,0)(120,0)(120,5)
\put(115,25){\line(1,0){20}}
\put(115,5){\line(0,1){20}}
\put(120,12){\mbox{$m$}}
\put(115,5){\line(1,0){20}}
\put(135,5){\line(0,1){20}}
\qbezier(130,25)(130,30)(135,30)
\qbezier(130,5)(130,0)(135,0)
\put(135,30){\vector(1,0){20}}
\put(135,0){\vector(1,0){20}}
\qbezier(155,-30)(160,-30)(160,-25)
\qbezier(155,0)(160,0)(160,-5)
\put(155,-25){\line(1,0){20}}
\put(155,-5){\line(0,-1){20}}
\put(157,-17){\mbox{$-m$}}
\put(155,-5){\line(1,0){20}}
\put(175,-5){\line(0,-1){20}}
\qbezier(170,-25)(170,-30)(175,-30)
\qbezier(170,-5)(170,0)(175,0)
\put(155,30){\line(1,0){40}}
\put(175,-30){\vector(1,0){60}}
\put(175,0){\vector(1,0){20}}
\qbezier(195,30)(200,30)(200,25)
\qbezier(195,0)(200,0)(200,5)
\put(195,25){\line(1,0){20}}
\put(195,5){\line(0,1){20}}
\put(200,12){\mbox{$m$}}
\put(195,5){\line(1,0){20}}
\put(215,5){\line(0,1){20}}
\qbezier(210,25)(210,30)(215,30)
\qbezier(210,5)(210,0)(215,0)
\put(215,30){\vector(1,0){60}}
\put(215,0){\vector(1,0){20}}
\qbezier(235,-30)(240,-30)(240,-25)
\qbezier(235,0)(240,0)(240,-5)
\put(235,-25){\line(1,0){20}}
\put(235,-5){\line(0,-1){20}}
\put(236,-17){\mbox{$-m$}}
\put(235,-5){\line(1,0){20}}
\put(255,-5){\line(0,-1){20}}
\qbezier(250,-25)(250,-30)(265,-30)
\qbezier(250,-5)(250,0)(255,0)
\put(255,-30){\vector(1,0){50}}
\put(255,0
){\vector(1,0){20}}

\qbezier(275,30)(280,30)(280,25)
\qbezier(275,0)(280,0)(280,5)
\put(275,25){\line(1,0){20}}
\put(275,5){\line(0,1){20}}
\put(280,12){\mbox{$m$}}
\put(330,0){\mbox{$\ldots n$}}
\put(275,5){\line(1,0){20}}
\put(295,5){\line(0,1){20}}
\qbezier(290,25)(290,30)(295,30)
\qbezier(290,5)(290,0)(295,0)
\put(295,-30){\vector(1,0){30}}
\put(295,0
){\vector(1,0){20}}
\put(295,30
){\vector(1,0){20}}

\end{picture}
\end{equation}
The classification of  knots belongs to the hybrid weaving knot  $\hat{W}_3(m,n)$ are tabulated below for some values of $m$ and $n$:
\begin{table}
\begin{center}
\begin{tabular} { | p {5 cm} | p {5 cm}| }
\hline
Notation & Knot \\
\hline
 $\hat{W}_3(1,n)$& weaving knot of type $W(3,n)$ \\
$\hat{W}_3(m,1)$& $T_{(2,m)}\#T^*_{(2,m)}$ \\
$\hat{W}_3(3,2)$ & $12a1288$ Knot \\
\hline
\end{tabular}
\caption{ The classification of hybrid weaving knot  $\hat{W}_3(m,n)$ }
\label{table1}
\end{center}
\end{table}
where $m$ is odd and $m\neq n>1$. When $m=1$,$\hat{W}_3(1, n)$ reduces to the  weaving knot $W(3,n)$ discussed in \cite{mishra2017jones, RRV}. Well known examples of weaving knots(see in Fig.\ref{HKD}) are $$ W(3,2)={\bf 4_1}, W(3,4)={\bf 8_{18}} ~\text{ and}~ W(3,5)={\bf 10_{123}}.$$ For $m>3$ and $n\geq 2$, the crossing number exceeds 20 whose data are not available in the knot theory literature to validate.  
Now we will elaborate the modified RT method for hybrid weaving knots and achieve a closed form expression for their HOMFLY-PT  polynomial.
\subsection{HOMFLY-PT for hybrid weaving knot $\hat{W}_3(m,n)$}
In this case, tensor product of fundamental representation of three strand braid: $$[1]^{\bigotimes 3}=[3]\bigoplus[1,1,1]\bigoplus 2 [2,1,0]$$ shows that representation $[2,1,0]$ has multiplicity two.  Incorporating $2\times 2$ matrix form ${\hat{\mathcal{R}}}_1$ and ${\hat{\mathcal{R}}}_2$ for representation $[2,1,0]$  in eqn.(\ref{HOMFLY}), the HOMFLY-PT for $\hat{W}_3(m,n)$ is
\begin{eqnarray}
\mathcal{H}_{[1]}^{\hat{W}_3(m,n)} &=& \frac{1}{S^*_{[1]}}\sum_{\Xi_{\alpha}=\{[3],[1,1,1],[2,1,0]\}}S^*_{\Xi_{\alpha}}{\rm Tr}_{\Xi_{\alpha}}  (\hat{\mathcal{R}}^{\Xi_{\alpha}}_1)^{m}(\hat{\mathcal{R}}^{\Xi_{\alpha}}_2)^{-m}\ldots (\hat{\mathcal{R}}^{\Xi_{\alpha}}_1)^{m}(\hat{\mathcal{R}}^{\Xi_{\alpha}}_2)^{-m}\nonumber\\
&=& \frac{1}{S^*_{[1]}} (S^*_{[3]}+ S^*_{[1,1,1]}+S^*_{[2,1]} {\rm Tr}_{[2,1,0]}  (\hat{\mathcal{R}}^{[2,1,0]}_1)^{m}(\hat{\mathcal{R}}^{[2,1,0]}_2)^{-m})^n{\label{HW}}~.
\end{eqnarray}
Here $S^*_{[1]}=[N]_q$, $S^*_{[3]}=\frac{[N]_{q}[N+1]_{q}[N+2]_{q}}{[2]_{q}[3]_{q}}$, $S^*_{[111]}=\frac{[N]_{q}[N-1]_{q}[N-2]_{q}}{[2]_{q}[3]_{q}}$, and $S^*_{[21]}=\frac{[N]_{q}[N+1]_{q}[N-1]_{q}}{[3]_{q}}$.\\
In order to apply the formula (\ref{HW}) to evaluate the HOMFLYPT polynomial for $\hat{W}_3(m,n)$ we need to compute the trace of the matrix $\Psi^{[2,1,0]}[m,n]=((\hat{\mathcal{R}}^{[2,1,0]}_1)^{m}(\hat{\mathcal{R}}^{[2,1,0]}_2)^{-m}))^n$. 
Using eqn(\ref{egn}) and eqn.(\ref{symrac}), we  have,
\begin{eqnarray*}
{\hat{\mathcal{R}}}_1&=& A^{-1}\left(\begin{array}{cc} q & 0 \\ \\  0 & -\frac{1}{q} \end{array}\right)~\text{and}~~ \hat{\mathcal{R}}_{2}
=A^{-1}\left(\begin{array}{cc} \frac{q^2 - [3]_q}{q {[2]_q}^2} &- \frac{\sqrt{[3]_q}}{[2]_{q}}\\ \\ - \frac{\sqrt{[3]_q}}{[2]_{q}}&  \frac{1-q^2 [3]_q}{q {[2]_q}^2}  \end{array}\right).\end{eqnarray*} 
Thus
\begin{eqnarray*}
 \hat{\mathcal{R}}^m_{1}{\hat{\mathcal{R}}_2}^{-m}
=\left(\begin{array}{cc} \frac{1-q^{2m}[3]_q}{([2]_q)^2} & -\frac{(1+q^{2 m})\sqrt{[3]_q}}{([2]_q)^2}\\ \\  \frac{(1+q^{2m})\sqrt{[3]_q}}{q^{2m}([2]_q)^2} & \frac{1-q^{-2m}[3]_q}{( [2]_q)^2} \end{array}\right)= \left(\begin{array}{cc} x_1 & -x_2\\ \\  \frac{x_2}{q^{2m}} & x_3 \end{array}\right),\end{eqnarray*}~
 where $x_1=\frac{1-q^{2m}[3]_q}{([2]_q)^2}$,  $x_2=\frac{(1+q^{2m})\sqrt{[3]_q}}{([2]_q)^2}$, and  $x_3=\frac{1-q^{-2m}[3]_q}{( [2]_q)^2}.$
Interestingly, we have succeed  in the writing of diagonal entries of  the $nth$ power of the above matrix $((\hat{\mathcal{R}}^{[2,1,0]}_1)^{m}(\hat{\mathcal{R}}^{[2,1,0]}_2)^{-m})^n$ in a compact form $\Psi^{[2,1,0]}_{1}[m,n]$ and $\Psi^{[2,1,0]}_{2}[m,n]$ i.e
\begin{eqnarray}{\label{traceterm}}
\Psi^{[2,1,0]}_{1}[m,n]&=&x_1^{n}+\sum_{i=1
}^{\floor{\frac{n}{2}}}\sum_{k=1}^{n-i}(-1)^i \binom{k+i-2}{i-1} \binom{n-(k+i-1)}{i}x_1^{n-(2i+k-1)} x_3^{k-1}(\frac{x_2}{q^m})^{2i}~,\nonumber\\
\Psi^{[2,1,0]}_{2}[m,n]&=&x_3^{n}+\sum_{i=1
}^{\floor{\frac{n}{2}}}\sum_{k=1}^{n-i}(-1)^i \binom{k+i-2}{i-1} \binom{n-(k+i-1)}{i}x_3^{n-(2i+k-1)} x_1^{k-1}(\frac{x_2}{q^m})^{2i}\nonumber\\.
\end{eqnarray}
Hence the trace of the matrix $\Psi^{[2,1,0]}[m,n]$
\begin{eqnarray}{\label{tt}}
\Psi^{[2,1,0]}[m,n]=\Psi^{[2,1,0]}_{1}[m,n]+\Psi^{[2,1,0]}_{2}[m,n].
\end{eqnarray}
 Using these binomial series for the trace,  the closed form  expression for HOMFLY-PT for hybrid weaving knot turns out to be 
\begin{eqnarray}{\label{HWK}}
\mathcal{H}_{[1]}^{\hat{W}_3 (m,n)}&=& \frac{1}{S^*_{[1]}}(S^*_{[3]}  + S^*_{[111]} + S^*_{[21]}\cdot( x_1^{n}+x_3^{n}+\sum_{i=1
}^{\floor{\frac{n}{2}}}\sum_{k=1}^{n-i}(-1)^i \binom{k+i-2}{i-1} \binom{n-(k+i-1)}{i}\nonumber\\&&(x_1^{(n+1-2i-k)} x_3^{(k-1)}+x_1^{(k-1)} x_3^{(n-2i-k+1)}))(\frac{x_2}{q^m})^{2i})).
\end{eqnarray}

The closed form expression is an important result providing a useful starting point to investigate $[r]$-colored HOMFLY-PT, knot-quiver correspondence for hybrid weaving knots which we will pursue in future.
Incidentally for $m=1$, $\Psi^{[2,1,0]}[m,n]$ in eqn(\ref{tt}) is a  Laurent polynomial \cite{RRV} giving closed 
form HOMFLY-PT for weaving knots $W(3,n)$. We propose such a Laurent polynomial structure will be seen for all  the  multiplicity two irreducible representation $\Xi_{\alpha}\in [r]^{\otimes 3}$  for symmetric colors $[r]>1$ as well.

\noindent
\textbf{Proposition 1.}  Given a representation $\Xi_{\alpha}\equiv[{\xi_1}^{\alpha},{\xi_2}^{\alpha},{\xi_3}^{\alpha}]$ having multiplicity 2  with  $\hat{\mathcal{R}}^{\Xi_{\alpha}}_1=\pm q^{m_1}A^{-r} \left(\begin{array}{cc} q^t & 0 \\ \\  0 & -\frac{1}{q^t} \end{array}\right)$, ~and~  $\mathcal{U}^{\Xi_{\alpha}}=\left(\begin{array}{cc}
\frac{1}{[2]_{q^t}} & \frac{\sqrt{[3]_{q^t}}}{ [2]_{q^t}} \\
\frac{\sqrt{[3]_{q^t}}}{[2]_{q^t}} & -\frac{1}{[2]_{q^t}}
\end{array}\right)\ \ $, the Laurent polynomial  $\mathcal{V}_{n,t}[q]$  is defined as 
\begin{eqnarray}{\label{conj}}
\mathcal{V}_{n,t}[q]&=&{\rm Tr}(\hat{\mathcal{R}}_1^{\Xi_{\alpha}} \mathcal{U}^{\Xi_{\alpha}}(\hat{\mathcal{R}}_1^{\Xi_{\alpha}})^{-1}(\mathcal{U}^{\Xi_{\alpha}})^{\dagger})^n=\sum_{g=-n}^{n} (-1)^{g} \mathcal{S}_{n,n-\abs{g}}q^{2 g t}.
\end{eqnarray}
Here  $t$ and $m_1$ are also an integer dependent on $\Xi_{\alpha}$ and the coefficients $\mathcal{S}_{n,j}$ are:
\begin{eqnarray*}
\mathcal{S}_{n,j}&=&\sum_{i=0}^{\floor{\frac{j}{2}}}\frac{n}{n-i}\binom{n-i}{n-j+ i} \binom{j-i-1}{i}~,
\end{eqnarray*}
where the parameters $n~\& j $ are positive integers and  $|x|$ denote the absolute value of $x$ and $\floor{x}$ indicate the greatest integer $x$. For $m=1$ and fundamental representation $[r]=[1]$, the trace in eqn.(\ref{tt}) is 
\begin{eqnarray*}
\Psi^{[2,1,0]}[1,n]=\mathcal{V}_{n,1}[q],
\end{eqnarray*}
exactly matching with the parallel work\cite{RRV}.
Further, we conjecture the sum of the absolute coefficient $\mathcal{S}_{n,n-|\delta|}$ given by $ \mathcal{O}_n$, satisfy the beautiful relation.
\vskip.1cm
\noindent
\textbf{Conjecture 1}:
\begin{eqnarray}{\label{febona}}
\mathcal{O}_n&=& \sum^{n}_{\delta=-n}\mathcal{S}_{n,n-\delta}=5\mathcal{F}_n^2+2 (-1)^{n}~,
 \end{eqnarray}
 \noindent
where $\mathcal{F}_n$ denotes Fibonacci numbers. The explicit form of $\mathcal{F}_{n}$ is given by \cite{pelitifibonacci}
\begin{eqnarray*}
\mathcal{F}_{n}&=&\frac{1}{\sqrt{5}}(\phi^{n}-\cos{(n \pi)}(\phi)^{-n})~.\\
\end{eqnarray*}
Here  $\phi \approx 1.618$ is the  golden ratio.
\noindent
We have checked this conjecture for large values of $n$. For values of $n\leq 8$, we have 
presented the values of  $\mathcal O_n, \mathcal F_n$ in Table.\ref{table2}. For completeness, we will briefly discuss the Fibonacci numbers and its properties.
The Fibonacci ($\mathcal{F}_n$) numbers are sequences satisfying the Fibonacci
recursion relation
\begin{eqnarray*}
\mathcal{F}_{n+1}=\mathcal{F}_n+\mathcal{F}_{n-1}, 
\end{eqnarray*}
with following initial conditions : $\mathcal{F}_0=0, \mathcal{F}_1=1$. Here $n$ is integer and it satisfy yhe following relation
\begin{eqnarray*}
\mathcal{F}_{-n}&=&(-1)^{n+1}\mathcal{F}_n.
\end{eqnarray*}
\begin{table} 
\centering
\small{
\begin{tabular} { | p {1 cm}| p {1 cm}| p {1 cm}| p {1 cm}| p {1 cm}| p {1 cm}| p {1 cm}| p {1 cm}| p {1 cm}|}
\hline
$n$ &1& 2& 3& 4& 5& 6& 7& 8\\
\hline
$\mathcal{F}_n$ &1&1& 2& 3& 5& 8& 13& 21\\
\hline
$\mathcal{O}_n$ &3&7& 18& 47& 123& 322& 843& 2207\\
\hline
\end{tabular}}
\caption{ $\mathcal O_n$ and $\mathcal F_n$ for $ n\leq 8$ }
\label{table2}
\end{table}
\subsection{Examples}
For the hybrid weaving knots in Table.\ref{table1}, HOMFLY-PT are obtained using our closed form expression for $\hat W_3(m,n)$.\\
$\bullet$ (a) For $m=1$,  the HOMFLY-PT polynomial is for weaving knots $W(3,n)$:
\begin{eqnarray}{\label{WEAVING}}
\mathcal{H}_{[1]}^{W(3,n)} (A,q)= \frac{1}{S^*_{[1]}}(S^*_{[3]}  + S^*_{[111]} + S^*_{[21]} \mathcal{V}_{n,1}[q]).
\end{eqnarray}
Substituting  $A=q^2$, we get the Jones polynomial:
\begin{eqnarray*}
\mathcal{J}^{W(3,n)}(q)=q^{-2}+q^2+\mathcal{V}_{n,1}[q]~.
\end{eqnarray*}
These results agree with the results in the  parallel paper on weaving knots\cite{RRV}.\\
$\bullet$ (b)  composite knot $T_{(2, m)}\#T^*_{(2, m)}$\\
For odd $m\geq 2$ and $n=1$, the knot belongs to composite knot of type $T_{(2, m)}\#T^*_{(2, m)}$\footnote{$T^*_{(2, m)}$ is the mirror of torus knot $T_{(2, m)}$ } . Hence, the HOMFLY-PT will be
\begin{eqnarray*}
H_{[1]}^{\hat{W}_3(m, 1)}&=&H_{[1]}^{T_{(2, m)}}(q,A) H_{[1]}^{T^*_{(2,m)}}(q,A))\\&&=
\frac{q^{2 - 2 m} (-1 + A^2 q^2 - A^2 q^{2 m} + q^{2 + 2 m}) (-A^2 + q^2 - 
   q^{2 m} + A^2 q^{2 + 2 m})}{A^2 (-1 + q)^2 (1 + q)^2 (1 + q^2)^2}.
\end{eqnarray*}

\begin{table}[h]
\begin{center}
\begin{tabular} { | p {2 cm} | p {2cm}|  p {8cm}|}
\hline
m &KNOT& $H_{[1]}^{\hat{W}_3(m, 1)}(q,A)$ \\
\hline
3 & $3_1\# 3^*_1$ & $(A^{-2} q^{-4})(1 - A^2 q^2 + q^4) (A^2 - q^2 + A^2 q^4)$ \\
\hline
9 &$9_1\# 9^*_1$&$ A^{-2} q^{-16}(1 - A^2 q^2 + q^4 - A^2 q^6 + q^8 - A^2 q^{10} + q^{12} - A^2 q^{14} + 
   q^{16})(A^2 - q^2 + A^2 q^4 - q^6 + A^2 q^8 - q^{10} + A^2 q^{12} - 
   q^{14} + A^2 q^{16})$\\
\hline
\end{tabular}
\caption{HOMFLY-PT for  $\hat{W}_3(m, 1)=T_{(2,m)}\# T^*_{(2,m)}$}\label{table3}
\end{center}
\end{table}

$\bullet$ (c) The $m=3$ and $n=2$ refers to a $12$ crossing knot $"12a1288"$ in the Rolfsen table whose HOMFLY-PT  polynomial is\\
\begin{eqnarray*}
H_{[1]}^{W_3(3, 2)}&=&11+\frac{7}{A^2}+7 A^2-\frac{1}{q^{10}}+\frac{1}{q^8}+\frac{1}{A^2 q^8}+\frac{A^2}{q^8}-\frac{5}{q^6}-\frac{1}{A^2 q^6}-\frac{A^2}{q^6}+\frac{6}{q^4}\\&&+\frac{4}{A^2
q^4}+\frac{4 A^2}{q^4}-\frac{11}{q^2}-\frac{5}{A^2 q^2}-\frac{5 A^2}{q^2}-11 q^2-\frac{5 q^2}{A^2}-5 A^2 q^2+6 q^4+\frac{4 q^4}{A^2}\\&&+4 A^2 q^4-5
q^6-\frac{q^6}{A^2}-A^2 q^6+
q^8+\frac{q^8}{A^2}+A^2 q^8-q^{10}.
\end{eqnarray*}
In the following section, we will present $[r]$-colored HOMFLYPT for $W(3,n)$ for $[r]=2,3$ 
and verify our proposition 1.
\section{Colored HOMFLY-PT for weaving knot type $W(3,n)$}\label{s4.int} 
We will use the data on $U^{\Xi_{\alpha}}$ matrices in section~\ref{s12.int} for three-strand braid where $\Xi_{\alpha} \in  [2]^3$ and $\Xi_{\alpha} \in [3]^3 $ to compute colored HOMFLY-PT for the weaving knots.
\subsection{Representation $[r]=[2]$\label{se.symrac1}}
In this case, $\bigotimes^3 [2] = [6,0,0]\bigoplus [3,3,0]\bigoplus  [4,1,1]\bigoplus 2 [5,1,0]\bigoplus 2 [3,2,1]\bigoplus 3 [4,2,0]$.\\
From the multiplicity, we can see that there  one $3\times 3$ matrix, two $2\times 2$ matrices, three $1\times 1$ matrices as shown in the Table.{\ref{table4}}.
{\small \begin{table}[h]
\begin{center}
\begin{tabular}{ |p{7cm}|p{2cm}|p{2cm}|  }
\hline
\hspace{3cm}$\Xi_{\alpha} \in [2]^3$   & Matrix size & $\# $ of matrices \\
\hline
[6,0,0], [4,1,1], [3,3]& 1 &3 \\
\hline
[5,1,0] ,[3,2,1]&2  & 2 \\
\hline
[4,2,0]& 1 & 3 \\
\hline
\end{tabular}
\caption{The multiplicity table for $\Xi_{\alpha} \in  [2]^3$ }\label{table4}
\end{center}
\end{table}
}
Also the path and the block structure of $\Xi \in [2]^{\otimes 3}$ is shown(\ref{2stand})
\begin{equation}{\label{2stand}}
\definecolor{ffqqtt}{rgb}{1.,0.,0.2}
\definecolor{ffqqqq}{rgb}{1.,0.,0.}
\definecolor{sqsqsq}{rgb}{0.12549019607843137,0.12549019607843137,0.12549019607843137}
\definecolor{qqzzff}{rgb}{0.,0.6,1.}
\begin{tikzpicture}[line cap=round,line join=round,>=triangle 45,x=.8
0cm,y=1cm,scale=.23]
\draw (9.335336945031727,14.0
49996666857155) node[anchor=north west] {$\mathbf{[4,0]}$};
\draw (23.551228417899944,14.496702093213744) node[anchor=north west] {$\mathbf{[3,1]}$};
\draw (38.50723616983793,14.496702093213744) node[anchor=north west] {$\mathbf{[2,2]}$};
\draw [color=sqsqsq](23.35180981324878,20.072864883911418) node[anchor=north west] {\textbf{[2]$\otimes$[2]}};
\draw [->,line width=1.pt,color=sqsqsq] (25.37790697674419,18.147653467420902) -- (10.562015503875973,14.150560444165093);
\draw [->,line width=1.pt,color=sqsqsq] (25.37790697674419,18.147653467420902) -- (25.377906976744196,14.257149591451917);
\draw (39.12700361169839,8.054647829647852) node[anchor=north west] {$\mathbf{[4,2]}$};
\draw (24.684174154334052,8.061236976934673) node[anchor=north west] {$\mathbf{[5,1]}$};
\draw (9.068864076814673,8.4600771860655603) node[anchor=north west] {$\mathbf{[6,0]}$};
\draw [color=sqsqsq](22.938050123326303,8.08640077709867) node[anchor=north west] {\textbf{2}};
\draw [color=sqsqsq](37.27429043340382,8.042050930423045) node[anchor=north west] {\textbf{3}};
\draw [->,line width=1.pt,color=sqsqsq] (25.37790697674419,18.09435889377752) -- (39.980620155038764,14.363738738738757);
\draw [->,line width=1.pt,color=sqsqsq] (25.377906976744185,12.76490152943643) -- (25.324612403100776,8.128273622459684);
\draw [->,line width=1.pt,color=sqsqsq] (11.041666666666666,12.445134087575966) -- (25.324612403100776,8.128273622459684);
\draw [->,line width=1.pt,color=sqsqsq] (10.455426356589147,12.338544940289143) -- (10.508720930232558,8.607924785250383);
\draw [->,line width=1.pt,color=sqsqsq] (11.319909418973037,12.36103833034904) -- (40.30038759689923,7.968389901529451);
\draw [->,line width=1.pt,color=sqsqsq] (26.017441860465116,12.71160695579302) -- (40.30038759689923,7.968389901529451);
\draw [->,line width=1.pt,color=sqsqsq] (40.35368217054264,12.65831238214961) -- (40.30038759689923,7.968389901529451);
\end{tikzpicture}
\end{equation}
The eigenvalues and $U^{\Xi_{\alpha}}$ matrices in this case are 
\begin{equation}\label{u0}
\hat{\mathcal{R}}^{[5,1,0]}_1=A^{-2}q^4\left(
\begin{array}{cc}
 q^2 & 0 \\
 0 & -q^{-2} \\
\end{array}
\right)\\,
\hat{\mathcal{R}}^{[3,2,1]}_1=A^{-2}q\left(
\begin{array}{cc}
 q^{-1} & 0 \\
 0 & -q \\
\end{array}
\right)\\,~~
\hat{\mathcal{R}}^{[4,2,0]}_1=A^{-2}\left(
\begin{array}{ccc}
 1 & 0 & 0 \\
 0 & -q^2 & 0 \\
 0 & 0 & q^6 \\
\end{array}
\right)
\end{equation}
\begin{equation}\label{u1}
\mathcal{U}^{[5,1,0]}=\left(\begin{array}{cc}
\frac{1}{[2]_{q^2}} & \frac{\sqrt{[3]_{q^2}}}{[2]_{q^2}} \\
\frac{\sqrt{[3]_{q^2}}}{[2]_{q^2}} & -\frac{1}{[2]_{q^2}}
\end{array}\right),\ \
~\mathcal{U}^{[3,2,1]}=\left(\begin{array}{cc}
\frac{1}{[2]_{q}} & \frac{\sqrt{[3]_{q}}}{[2]_{q}} \\
\frac{\sqrt{[3]_{q}}}{[2]_{q}} & -\frac{1}{[2]_{q}}
\end{array}\right),\ \
\end{equation}

\begin{equation}{\label{u2}}
\mathcal{U}^{[4,2,0]}=\left(
\begin{array}{ccc}
 -\frac{1}{1+\frac{1}{q^2}+q^2} & -\frac{q}{\sqrt{1+q^2+q^4}} & -\frac{\sqrt{1+q^2+q^4+q^6+q^8}}{1+q^2+q^4} \\
 -\frac{q}{\sqrt{1+q^2+q^4}} & -1+\frac{q^2}{1+q^4} & \frac{q^4 \sqrt{\left(1+\frac{1}{q^2}+q^2\right) \left(1+\frac{1}{q^4}+\frac{1}{q^2}+q^2+q^4\right)}}{\left(1+q^4\right)
\left(1+q^2+q^4\right)} \\
 -\frac{\sqrt{1+q^2+q^4+q^6+q^8}}{1+q^2+q^4} & \frac{q^4 \sqrt{\left(1+\frac{1}{q^2}+q^2\right) \left(1+\frac{1}{q^4}+\frac{1}{q^2}+q^2+q^4\right)}}{\left(1+q^4\right)
\left(1+q^2+q^4\right)} & -\frac{q^4}{1+q^2+2 q^4+q^6+q^8} \\
\end{array}
\right)
\end{equation}
From  eqn.(\ref{HOMFLY}), $[2]$-HOMFLY-PT for $W(3,n)$:
\begin{eqnarray}
\mathcal{H}_{[2]}^{W(3,n)} &=& \frac{1}{S^*_{[2]}}\sum_{\alpha} S^*_{\Xi_{\alpha}} {\rm Tr}_{\Xi_{\alpha}} (\hat{\mathcal{R}}^{\Xi_{\alpha}}_1)(\hat{\mathcal{R}}^{\Xi_{\alpha}}_2)^{-1}\ldots (\hat{\mathcal{R}}^{\Xi_{\alpha}}_1)(\hat{\mathcal{R}}^{\Xi_{\alpha}}_2)^{-1}\nonumber\\
&=& \frac{1}{S^*_{[2]}}\sum_{\alpha}S^*_{\Xi_{\alpha}} {\rm Tr}_{\Xi_{\alpha}}  (\hat{\mathcal{R}}^{\Xi_{\alpha}}_1)(\hat{\mathcal{R}}^{\Xi_{\alpha}}_2)^{-1})^n\nonumber\\&=& \frac{1}{S^*_{[2]}} (S^*_{[6]}+ S^*_{[3,3]}+S^*_{[4,1,1]} +S^*_{[5,1,0]}{\rm Tr}_{[5,1,0]}  (\hat{\mathcal{R}}^{[5,1,0]})(\hat{\mathcal{R}}^{[5,1,0]}_2)^{-1})^n+S^*_{[3,2,1]}*\nonumber\\&&{\rm Tr}_{[3,2,1]}  (\hat{\mathcal{R}}^{[3,2,1]})(\hat{\mathcal{R}}^{[3,2,1]}_2)^{-1})^n+S^*_{[4,2,0]}{\rm Tr}_{[4,2,0]}  (\hat{\mathcal{R}}^{[4,2,0]})(\hat{\mathcal{R}}^{[4,2,0]}_2)^{-1})^n{\label{HW22}}.
\end{eqnarray}
Using eqns.$(\ref{u0})$ to $(\ref{u2})$, and $(\ref{conj})$, we can rewrite  the equation~(\ref{HW22}) into neat formula
\begin{eqnarray}
\mathcal{H}_{[2]}^{W{(3,n)}} &=& \frac{1}{S^*_{[2]}} (S^*_{[6]}+ S^*_{[3,3]}+S^*_{[4,1,1]} +S^*_{[5,1,0]} \mathcal{V}_{n,2}[q]+S^*_{[3,2,1]}\mathcal{V}_{n,1}[q]+S^*_{[4,2,0]}{\rm Tr} (X^{[4,2,0]})^n\nonumber{\label{HW2}}~,
\end{eqnarray}
where,
\begin{equation}
X^{[4,2,0]}=\left(
\begin{array}{ccc}
 \frac{1}{q^6+q^8+q^{10}} & -\frac{1}{q^5 \sqrt{1+q^2+q^4}} & \frac{\sqrt{1+q^2+q^4+q^6+q^8}}{q^2+q^4+q^6} \\
 \frac{1}{q^3 \sqrt{1+q^2+q^4}} & \frac{-1+q^2-q^4}{q^2+q^6} & -\frac{q^3 \sqrt{\frac{1+q^2+q^4+q^6+q^8}{1+q^2+q^4}}}{1+q^4} \\
 \frac{q^4 \sqrt{1+q^2+q^4+q^6+q^8}}{1+q^2+q^4} & \frac{q^7 \sqrt{\frac{1+q^2+q^4+q^6+q^8}{1+q^2+q^4}}}{1+q^4} & \frac{q^{14}}{1+q^2+2 q^4+q^6+q^8}
\\
\end{array}
\right).
\end{equation}
 Using eqn.(\ref{conj}), the [2]-colored reduced HOMFLY-PT polynomials for $W(3,n)$. 
 We would like to emphasize that the polynomial form of this algebraic expression for arbitrary $n$ is easily computable. We have  listed $[2]$ colored HOMFLY-PT in Appendix~\ref{app1} for some weaving knots.
\subsection{Representation [3]\label{t.symrac1}}
In this case, $\bigotimes^3 [3] = [9,0,0]\bigoplus [7,1,1]\bigoplus  [5,2,2]\bigoplus  [4,4,1]\bigoplus  [3,3,3]\bigoplus 3 [8,1,0]\bigoplus \\ 2[4,3,2]\bigoplus 2 [6,2,1]\bigoplus 2 [5,4,0]\bigoplus 3 [7,2,0]\bigoplus 2 [5,3,1]\bigoplus 4 [6,3,0],\\$
Thus, there are two $3\times 3$ matrices, four $2\times 2$ matrices, five $1\times 1$ matrices and one $4\times 4$ matrix tabulated below.
{\small
\begin{table}[h]
\begin{center}
\begin{tabular}{ |p{7cm}|p{2cm}|p{2cm}|  }
\hline
\hspace{3cm}$\Xi_{\alpha} \in [3]^3 $   & Matrix size & $\# $ of matrices \\
\hline
[9,0,0], [7,1,1], [5,2,2],[4,4,1],[3,3,3]& 1 &5 \\
\hline
[4,3,2],[6,2,1], [5,4,0],[5,3,1]& 2  & 4 \\
\hline
[8,1,0], [7,2,0]& 3 & 2 \\
\hline
[6,3,0]&4 & 1 \\
\hline
\end{tabular}
\caption{The multiplicity table for $\Xi_{\alpha} \in  [3]^3$ }
\end{center}
\end{table}
}
The braiding and $U^{\Xi_{\alpha}}$ matrices in this case are 

\begin{equation}{\label{col3}}
~{\hat{\mathcal{R}}^{[6,3,0]}=A^{-3}\left(
\begin{array}{cccc}
 -q^3 & 0 & 0 & 0 \\
 0 & q^5 & 0 & 0 \\
 0 & 0 & -q^9 & 0 \\
 0 & 0 & 0 & q^{15} \\
\end{array}
\right)}\\,~~{\hat{\mathcal{R}}^{[5,3,1]}=A^{-3}\left(
\begin{array}{ccc}
 -q^3 & 0 & 0 \\
 0 & q^5 & 0 \\
 0 & 0 & -q^9 \\
\end{array}
\right)}\\,~~{\hat{\mathcal{R}}^{[7,2,0]}=A^{-3}\left(
\begin{array}{ccc}
 q^5 & 0 & 0 \\
 0 & -q^9 & 0 \\
 0 & 0 & q^{15} \\
\end{array}
\right)}~,
\end{equation}

\begin{equation}{\label{col31}}
\hat{\mathcal{R}}^{[5,4,0]}=\hat{\mathcal{R}}^{[6,2,1]}=A^{-3}q^7\left(
\begin{array}{cc}
 q^{-2} & 0 \\
 0 & -q^2 \\
\end{array}
\right)\\,~\hat{\mathcal{R}}^{[4,3,2]}=A^{-3}q^4\left(
\begin{array}{cc}
 -q^{-1} & 0 \\
 0 & q \\
\end{array}
\right) \\,~
~{\hat{\mathcal{R}}^{[8,1,0]}=A^{-3}q^{12}\left(
\begin{array}{cc}
 -q^{-3} & 0 \\
 0 & q^{3} \\
\end{array}
\right)},
\end{equation}~

\begin{equation}{\label{col32}}
\mathcal{U}^{[5,3,1]}=\left(
\begin{array}{ccc}
 -\frac{1}{1+\frac{1}{q^2}+q^2} & -\frac{q}{\sqrt{1+q^2+q^4}} & -\frac{\sqrt{1+q^2+q^4+q^6+q^8}}{1+q^2+q^4} \\
 -\frac{q}{\sqrt{1+q^2+q^4}} & -1+\frac{q^2}{1+q^4} & \frac{q^4 \sqrt{\left(1+\frac{1}{q^2}+q^2\right) \left(1+\frac{1}{q^4}+\frac{1}{q^2}+q^2+q^4\right)}}{\left(1+q^4\right)
\left(1+q^2+q^4\right)} \\
 -\frac{\sqrt{1+q^2+q^4+q^6+q^8}}{1+q^2+q^4} & \frac{q^4 \sqrt{\left(1+\frac{1}{q^2}+q^2\right) \left(1+\frac{1}{q^4}+\frac{1}{q^2}+q^2+q^4\right)}}{\left(1+q^4\right)
\left(1+q^2+q^4\right)} & -\frac{q^4}{1+q^2+2 q^4+q^6+q^8} \\
\end{array}
\right)~,
\end{equation}
\begin{equation}{\label{col33}}
\mathcal{U}^{[5,4,0]}=\mathcal{U}^{[3,6,2,1]}=\left(\begin{array}{cc}
\frac{1}{[2]_{q^2}} & \frac{\sqrt{[3]_{q^2}}}{[2]_{q^2}} \\
\frac{\sqrt{[3]_{q^2}}}{[2]_{q^2}} & -\frac{1}{[2]_{q^2}}
\end{array}\right),\ \
\end{equation}
\begin{equation}{\label{col34}}
\mathcal{U}^{[8,1,0]}=\left(\begin{array}{cc}
\frac{1}{[2]_{q^3}} & \frac{\sqrt{[3]_{q^3}}}{[2]_{q^3}} \\
\frac{\sqrt{[3]_{q^3}}}{[2]_{q^3}} & -\frac{1}{[2]_{q^3}}
\end{array}\right),\ \
\mathcal{U}^{[4,3,2]}=\left(\begin{array}{cc}
\frac{1}{[2]_{q}} & \frac{\sqrt{[3]_{q}}}{[2]_{q}} \\
\frac{\sqrt{[3]_{q}}}{[2]_{q}} & -\frac{1}{[2]_{q}}
\end{array}\right),\ \
\end{equation}
We have placed the other $3 \times 3$ and also $4\times 4$  matrices in Appendix~\ref{app}.
From  eqn.(\ref{HOMFLY}), $[3]$-colored  HOMFLY-PT for $W(3,n)$:
\begin{eqnarray*}
\mathcal{H}_{[3]}^{W(3,n)} &=& \frac{1}{S^*_{[3]}}\sum_{\alpha} S^*_{\Xi_{\alpha}} {\rm Tr}_{\Xi_{\alpha}} (\hat{\mathcal{R}}^{\Xi_{\alpha}}_1)(\hat{\mathcal{R}}^{\Xi_{\alpha}}_2)^{-1}\ldots (\hat{\mathcal{R}}^{\Xi_{\alpha}}_1)(\hat{\mathcal{R}}^{\Xi_{\alpha}}_2)^{-1}~,\nonumber\\&=& \frac{1}{S^*_{[3]}}\sum_{\alpha}S^*_{\Xi_{\alpha}} {\rm Tr}_{\Xi_{\alpha}}  (\hat{\mathcal{R}}^{\Xi_{\alpha}}_1)(\hat{\mathcal{R}}^{\Xi_{\alpha}}_2)^{-1})^n~,\nonumber\\&=& \frac{1}{S^*_{[3]}} (S^*_{[9]}+ S^*_{[7,1,1]}+S^*_{[5,2,2]} +S^*_{[4,4,1]}+S^*_{[3,3,3]}+S^*_{[4,3,2]}{\rm Tr}_{[4,3,2]}  (\hat{\mathcal{R}}^{[4,3,2]})(\hat{\mathcal{R}}^{[4,3,2]}_2)^{-1})^n+\\&&S^*_{[6,2,1]}{\rm Tr}_{[6,2,1]}  (\hat{\mathcal{R}}^{[6,2,1]})(\hat{\mathcal{R}}^{[6,2,1]}_2)^{-1})^n+S^*_{[5,4,0]}{\rm Tr}_{[5,4,0]}  (\hat{\mathcal{R}}^{[5,4,0]})(\hat{\mathcal{R}}^{[5,4,0]}_2)^{-1})^n+\\&&S^*_{[8,1,0]}{\rm Tr}_{[8,1,0]}  (\hat{\mathcal{R}}^{[8,1,0]})(\hat{\mathcal{R}}^{[8,1,0]}_2)^{-1})^n+S^*_{[7,2,0]}{\rm Tr}_{[7,2,0]}  (\hat{\mathcal{R}}^{[7,2,0]})(\hat{\mathcal{R}}^{[7,2,0]}_2)^{-1})^n+\\&&S^*_{[6,3,0]}{\rm Tr}_{[6,3,0]}  (\hat{\mathcal{R}}^{[6,3,0]})(\hat{\mathcal{R}}^{[6,3,0]}_2)^{-1})^n{\label{HW3}}.
\end{eqnarray*}
Using eqns.($\ref{col3})$ to $(\ref{col34})$, eqn.(\ref{conj}), and Appendix~\ref{app}, we can rewrite  the equation(\ref{HW3}) into neat formula
\begin{eqnarray}
\mathcal{H}_{[3]}^{W{(3,n)}} &=& \frac{1}{S^*_{[3]}} (S^*_{[9]}+ S^*_{[7,1,1]}+S^*_{[5,2,2]} +S^*_{[4,4,1]}+S^*_{[3,3,3]} +(S^*_{[6,2,1]}+S^*_{[5,4,0]}) \mathcal{V}_{n,2}[q]\nonumber\\&&+S^*_{[4,3,2]}\mathcal{V}_{n,1}[q]+S^*_{[8,1,0]}\mathcal{V}_{n,3}[q]+S^*_{[7,2,0]}{\rm Tr} (X_1^{[7,2,0]})^n+S^*_{[5,3,1]}{\rm Tr} (X_2^{[5,3,1]})^n\nonumber\\&&
+S^*_{[6,3,0]}{\rm Tr} (X_3^{[6,3,0]})^n{\label{HW33}}.\end{eqnarray}
where the explicit form of $X_3^{[6,3,0]}$, $X_1^{[7,2,0]}$ and $X_2^{[5,3,1]}$ are given in Appendix \ref{app} and the colored HOMFLY-PT polynomials for $W(3,n)$ for color [3] are presented in Appendix~\ref{app1}.
Even though we have explicitly computed $[r]$-colored HOMFLY-PT upto $[r]=3$, the method is straightforward. However, it will be interesting if we can write a closed form expression for arbitrary color $[r]$. This is essential to work on volume conjecture for these hyperbolic knots which we plan to pursue in future. As a piece of evidence that our $[r]$-colored HOMFLY-PT for weaving knots are correct, we work out reformulated invariants and BPS integers in the context of topological string duality in the following section.
\section{Integrality structures in topological strings }\label{s5.int}
Motivated by the AdS-CFT correspondence, Gopakumar-Vafa conjectured that the $SU(N)$  Chern-Simons theory on $S^3$ is dual to closed A-model topological string theory on a resolved conifold $\cal{O}$(-1)  + $\cal{O}$ (-1) over $\mathbf P^1$. Particularly, the Chern-Simons free energy $\ln Z[S^3]$ was shown to be closed string partition function on the resolved conifold target space:
\begin{equation}
\ln Z[S^3]=-\sum_g \mathcal{F}_{g}(t) g_{s}^{2-2g},
\end{equation}
 where $\mathcal{F}_{g}(t)$ are the genus $g$ topological string amplitude, $g_s=\frac{2 \pi}{k+N}$ denotes the string coupling constant and $t=\frac{2 \pi i N}{k+N}$ denote the ${\rm K\ddot{a}hler}$ parameter of $\mathbf P^1$.  Ooguri-Vafa conjectured that the Wilson loop operators in Chern-Simons theory correspond to  the following topological string operator on a deformed conifold $T^{*}S^3$:
\begin{eqnarray}\label{uov}
\ln Z(U,V)_{\mathcal{S}^{3}}&=&\sum_{m}\frac{1}{m}Tr_{[1]} U^mTr_{[1]} V^m,
\end{eqnarray}
where $U$ represent the holonomy of the gauge connection $A$ around the knot $\mathcal{K}$ carrying
the fundamental representation($[1]$) in the $U(N)$ Chern-Simons theory on $S^3$, and V is the
holonomy of a gauge field $\tilde{A}$ around the same component knot carrying the fundamental
representation($[1]$) in the $U(M)$ Chern-Simons theory on a Lagrangian sub-manifold $\cal C$ which intersects $S^3$ along the knot $\mathcal{K}$. Gopakumar-Vafa duality require integrating the gauge field $A$ on $S^3$ leading to open topological string amplitude on the resolved conifold background. For unknot, the detailed calculation was performed \cite{OV} giving: 
\begin{eqnarray}
\left\langle Z(U,V)\right\rangle_{S^3}&=&\exp{(i \sum_{m=1}^{\infty} \frac{\exp {(\frac{m t}{2})}-\exp {(\frac{-m t}{2})}}{2 m \sin{(\frac{m g_s}{2})}}Tr V^{-m})},\label{gv}
\end{eqnarray}
which was justified using Gopakumar-Vafa duality. Further, Ooguri-Vafa conjectured the generalization of eqn.(~\ref{gv}) for other knots as (also known ${\bf LMOV}$ integrality conjecture):
\begin{eqnarray}
\left\langle Z(U,V)\right\rangle_{S^3}&=&\sum_{{\bf R}} \mathcal{H}^{*\mathcal{K}}_{{\bf R}}(q,{\bf A})  Tr_{{\bf R}} V\nonumber\\
&=&\exp\bigg[\sum_{m=1}^{\infty} \left(\sum\limits_ {R}\frac{1}{m} f^{\mathcal{K}}_{{\bf R}}({A}^m,q^m)Tr_{{\bf R}} V^m\right)\bigg],\label{guv}
\end{eqnarray}

 where $f^{\mathcal{K}}_{{\bf R}}({A},q)$, known as reformulated invariant, obeying the following integrality structure:
\begin{eqnarray}
f^{\mathcal{K}}_{{\bf R}} (q,  A)&=&\sum_{i,j} \frac{1}{(q-q^{-1})}{\widetilde {\bf N}^{\mathcal{K}}_{{\bf R},i,j}{A}^i q^{j}}\nonumber.
\end{eqnarray}
Here, $R$ denotes the irreducible representation of $U(N)$ and ${\widetilde {\bf N}}^{\mathcal{K}}_{{\bf R}, i,j}$ counts the number of D2-brane intersecting D4-brane (BPS states) where, i and j keeps track of charges and spins respectively\cite{GV1, GV2}. These reformulated invariants can be written in the terms of colored HOMFLY-PT polynomials \ref{guv}. For few lower dimensional representations, the explicit forms are as follows\cite{LMV, LM1, LM2}:
{
\begin{eqnarray} 
f^{\mathcal{K}}_{[1]}(q,{ A})&=&\mathcal{H}^{*\mathcal{K}}_{[1]}(q,{A}),\nonumber \\
f^{\mathcal{K}}_{[2]}(q,{ A})&=&\mathcal{H}^{*\mathcal{K}}_{[2]}(q,{ A})-{1\over 2}\Big(\mathcal{H}^{*\mathcal{K}}_{[1]}(q,{ A})^2+\mathcal{H}^{*\mathcal{K}}_{[1]}(q^2,{A}^2)\Big),\nonumber\\
f^{\mathcal{K}}_{[1^2]}(q,{ A})&=&\mathcal{H}^{*\mathcal{K}}_{[1^2]}(q,A)-{1\over 2}\Big( \mathcal{H}^{*\mathcal{K}}_{[1]}(q,A)^2-\mathcal{H}^{*\mathcal{K}}_{[1]}(q^2,A^2)\Big),\nonumber\\
\ldots \nonumber
\end{eqnarray}}
In fact, reformulated invariants obey Ooguri-Vafa conjecture verified for many arborescent knots up to 10 crossings in \cite{Mironov:2017hde}. Moreover, these reformulated invariant can be equivalently written as \cite{LMV}:
\begin{eqnarray}\label{ic}
f^{\mathcal{K}}_{{\bf R}} (q,{ A})=\sum_{m,k\ge 0,s} C_{{\bf R}{\bf S}}\hat {\bf N}^{\mathcal{K}}_{{\bf S},m,k}{ A}^m(q-q^{-1})^{2k-1},
\end{eqnarray}
where $\hat {\bf N}^{\mathcal{K}}_{{\bf S},m,k}$ called refined integers and
\begin{eqnarray}
C_{{\bf R}{\bf S}}={1\over q-q^{-1}}\sum_{\Delta}{1\over z_\Delta}\psi_{{\bf R}}(\Delta)\psi_{{\bf S}}(\Delta){\prod_{i=1}^{l(\Delta)}\Big(q^{\xi_i}-q^{-\xi_i}\Big)}\nonumber.
\end{eqnarray}
 Here the sum goes over the Young diagrams $\Delta$ with $l(\Delta)$ lines of lengths $\xi_i$ and the number of boxes $|\Delta|=\sum_{i}^{l(\Delta)} \xi_i$, while  $\psi_{{\bf R}}(\Delta)$ denote the characters of symmetric groups at $|R|=|\Delta|$ and $z_{\Delta}$ is the standard symmetric factor of the Young diagram\cite{Fulton}. 
Using our colored HOMFLY-PT form for the weaving knot $W(3,n)$ (listed in Appendix ~\ref{app1}), we computed the reformulated invariants for representations upto length  $|{\bf R}|=2$.  From our  analysis, we propose the following:\\
\vskip1cm
\begin{mdframed}[style=sid]
\textbf{Proposition 2.} \emph{Refined BPS integer $\hat{\bf N}^{W(3,n)}_{[ 1],\mp1,k}$ for weaving knot $W(3,n)$  is the coefficient of $z^k$ of polynomial $f^{\mp}_{n}[z]$ of degree $n-1$ i.e
 \begin{eqnarray}{\label{CHP}}
f^{\mp}_{n}[z]&=& \pm \frac{2(-1)^n T_{n}(\frac{1+z}{2})+1}{z}~,
\end{eqnarray}}
\end{mdframed}
where $T_{n}(z)$  represents the $nth$ degree Chebyshev polynomial of the first kind at the point z. Rodrigue's formula to obtain $T_n(z)$ is 
\begin{equation}
T_n(z) = \frac{(-2)^n n!}{2n!} \sqrt{1-z^2} \frac{d^n} {dz^n} (1-z^2)^{n-1/2}~.
\end{equation}
Here we list the polynomial form for some values of $n$:
For completeness,
\begin{eqnarray*}
f^{-}_{11}[z]&=&11 + 22 z - 66 z^2 - 99 z^3 + 77 z^4 + 154 z^5 + 22 z^6 - 66 z^7 - 
 44 z^8 - 11 z^9 - z^{10}\\
f^{-}_{10}[z]&=&-10 + 15 z + 60 z^2 - 15 z^3 - 98 z^4 - 35 z^5 + 40 z^6 + 
 35 z^7 + 10 z^8 + z^{9}\\
f^{-}_{5}[z]&=&5 + 5 z - 5 z^2 - 5 z^3 - z^4\\
f^{-}_{4}[z]&=&-4 + 2 z + 4 z^2 + z^3
\end{eqnarray*}
Unfortunately, we have not managed to write the other integers for fundamental representation $\hat{\bf N}^{W(3,n)}_{[ 1],\pm 3,k}:$ as a closed form. There are other properties of $\hat{\bf N}$ which we have checked 
 up to the level $|{\bf S}|=2$ for $W(3,n)$ knot. They are    
\begin{eqnarray*}
\sum_m \hat{\bf N}^{W(3,n)}_{{\bf S},m,k}&=0&\\
\sum_k \hat{\bf N}^{W(3,n)}_{{\bf [1]
},\mp 1 ,k}&=&\mp \frac{4}{3}  T_{n-1}(-1)\sec(\frac{n \pi}{6})^2 \sin(\frac{n \pi}{3})^4 , ~ n\geq 1.
\end{eqnarray*}
here $T_{n-1}(z=-1)$ is the Chebyshev polynomial evaluated at $z=-1$.
We have tabulated below these
refined integers for knot $W(3,4), W(3,5), W(3,10)$, and $W(3,11)$, when $|r=1|$:
\begin{center}\begin{tabular}{cccc}
$\hat{\bf N}^{W[3,4]}_{[ 1]}:$ &
\begin{tabular}{|c|cccc|}
\hline
&&&&\\
$ k \backslash m=$ & -3 & -1 & 1 & 3 \\
&&&&\\
\hline
&&&&\\
0 & 1 & -4 &4 & -1 \\
&&&&\\
1 & -1 &2 & -2 & 1 \\
&&&&\\
2 & -1 & 4 & -4 & 1 \\
&&&&\\
4 & 0 & 1 & -1 & 0 \\
&&&&\\
\hline
\end{tabular}
\end{tabular}, \begin{tabular}{cccc}
$\hat{\bf N}^{W[3,5]}_{[ 1]}:$ &
\begin{tabular}{|c|cccc|}
\hline
&&&&\\
$ k \backslash m=$ & -3 & -1 & 1 & 3 \\
&&&&\\
\hline
&&&&\\
0 & -2 & 5 &- 5 & 2 \\
&&&&\\
1 & -1 & 5 & -5 & +1 \\
&&&&\\
2 & 2 & -5 & 5 & -2 \\
&&&&\\
3 & 1 & -5 & 5 & -1 \\
&&&&\\
4 & 0 & -1 & 1 & 0 \\
&&&&\\
\hline
\end{tabular}
\end{tabular}
\end{center}
\begin{center}\begin{tabular}{cccc}
$\hat{\bf N}^{W[3,10]}_{[ 1]}$ &
\begin{tabular}{|c|cccc|}
\hline
&&&&\\
$ k \backslash m=$ & -3 & -1 & 1 & 3 \\
&&&&\\
\hline
&&&&\\
0 & 3 & -10 &10 & 3\\
&&&&\\
1 & -6& 15 & -15 & 6 \\
&&&&\\
2 & -18 & 60 & -60 & 18 \\
&&&&\\
3 & 11 & -15 & 15 & -11 \\
&&&&\\
4 & 29 & -98 & 98 & -29 \\
&&&&\\
5 & 2 & -35 & 35 & -2 \\
&&&&\\
6 & -14 & 40 & -40 & 14 \\
&&&&\\
7 & -7 & 35 &-35 &7 \\
&&&&\\
8 & -1 & 10 & -10 & 1\\
&&&&\\
9 & 0 & 1 & -1 & 0 \\
&&&&\\

\hline
\end{tabular}
\end{tabular},~\begin{tabular}{cccc}
$\hat{\bf N}^{W[3,11]}_{[ 1]}$ &
\begin{tabular}{|c|cccc|}
\hline
&&&&\\
$ k \backslash m=$ & -3 & -1 & 1 & 3 \\
&&&&\\
\hline
&&&&\\
 0&-4 & 11 & -11 & 4 \\
&&&&\\
 1&-6 & 22 & -22 & 6 \\
&&&&\\
 2&24 & -66 & 66 & -24 \\
&&&&\\
3& 25 & -99 & 99 & -25 \\
&&&&\\
4& -34 & 77 & -77 & 34 \\
&&&&\\
5& -40 & 154 & -154 & 40 \\
&&&&\\
6& 6 & 22 & -22 & -6 \\
&&&&\\
7& 20 & -66 & 66 & -20 \\
&&&&\\
8& 8 & -44 & 44 & -8 \\
&&&&\\
9& 1 & -11 & 11 & -1 \\
&&&&\\
10& 0 & -1 & 1 & 0 \\
&&&&\\
\hline
\end{tabular}
\end{tabular}
\end{center}

The table of refined integers for representations  whose length $|R|= 2$  are presented in Appendix ~\ref{app2}.
\section{Conclusion and discussion}
\label{s6.int}
Hybrid weaving knots $\hat W_3(m,n)$ obtained  from  braid word $\left[\sigma_1^m \sigma_2^{-m}\right]^n$ (see Fig.\ref{GWK})  contains weaving knots $W(3,n)$ as subset which are hyperbolic in nature. Finding a closed form expression for $[r]$-colored HOMFLY-PT for such hybrid weaving knots was attempted using the modified Reshtikhin-Turaev approach \cite{RT1}-\cite{RT2} method. 
Using the $\hat{\mathcal{R}_i}$ matrices,  we derived the explicit closed form expression of HOMFLY-PT for hybrid weaving knot$\hat{W}_3(m,n)$ (\ref{HWK}). Motivated by the Laurent polynomial structure studied for HOMFLY-PT of weaving knots\cite{RRV}, we proposed such a structure $\mathcal{V}_{n,t}[q]$ (\ref{HW22} and \ref{HW33}) for any $[r]$-colored HOMFLY-PT for the weaving knots. Further we showed that the absolute sum of the coefficients in the Laurent polynomial is related to Fibonacci numbers (see conjecture 1 (\ref{febona})). We have  computed the colored HOMFLY-PT for $W(3,n)$ upto $[r]=3$ and presented them in the appendix~\ref{app1}. Clearly, writing the polynomial form is computationally simplified by this modified RT method.  Using these knot invariants, we computed reformulated invariants and found some of the refined BPS integers can be written in terms of coefficient of Chebyshev polynomials($T_{n}(x))$ of first kind for $W(3,n)$ (\ref{CHP}).\\
So far, we have have managed to write the closed form expression for trace of 2x2 matrices by introducing the $\mathcal{V}_{n,t}[q]$. For higher dimensional matrices, such a Laurent polynomial structure is not obvious.  We have seen  a concise  form for $[r]$-colored HOMFLY-PT  for knot $4_1\equiv W(3,2)$, twist knots and torus knots  using $q$-binomial and $q$-Pochammer  terms.\\ 
It will be interesting if we can find a similar expression for weaving knots. Such an expression will help us to address volume conjecture, A-polynomials for these weaving knots. We hope to address these problems in future. 

\vspace{0.5cm}

\textbf{Acknowledgements}
 VKS would like to acknowledge the hospitality of department of mathematics, IISER, Pune (India) where this work was done during his visit as visiting fellow. PR would like to thank SERB ((MATRICS) MTR/2019/000956 funding. 

\bibliographystyle{JHEP}
\bibliography{HWF}

\section{Appendix A}{\label{app}}

\begin{equation}
\mathcal{U}^{[6,3,0]}=\left(
\begin{array}{cccc}
 \frac{q^3}{1+q^2+q^4+q^6} & \frac{q^2 \sqrt{1+q^2+q^4}}{1+q^2+q^4+q^6} &  (u_1)_{13}& (u_1)_{14}
\\
 \frac{q^2 \sqrt{1+q^2+q^4}}{1+q^2+q^4+q^6} &\frac{q+2 q^3+2 q^5+q^7+2 q^9+2 q^{11}+q^{13}}{\left(1+q^2+q^4+q^6\right) \left(1+q^2+q^4+q^6+q^8\right)}
& (u_1)_{23} &  \\
 \frac{q \sqrt{1+q^2+q^4+q^6+q^8}}{1+q^2+q^4+q^6} & (u_1)_{32} & -\frac{q+q^3-q^5+q^7+q^9}{1+q^4+q^6+q^{10}} &(u_1)_{34}  \\
 \frac{\sqrt{1+q^2+q^4+q^6+q^8+q^{10}+q^{12}}}{1+q^2+q^4+q^6} &(u_1)_{42} &(u_1)_{43}& (u_1)_{44}\\
\end{array}
\right)
\end{equation}
\begin{eqnarray*}
(u_1)_{14}&=& \frac{\sqrt{1+q^2+q^4+q^6+q^8+q^{10}+q^{12}}}{1+q^2+q^4+q^6}\\
(u_1)_{13}&=& \frac{q \sqrt{1+q^2+q^4+q^6+q^8}}{1+q^2+q^4+q^6}\\
(u_1)_{24}&=&-\frac{q^5 \left(1+q^2+q^4\right) \sqrt{\left(1+\frac{1}{q^2}+q^2\right) \left(1+\frac{1}{q^6}+\frac{1}{q^4}+\frac{1}{q^2}+q^2+q^4+q^6\right)}}{\left(1+q^2+q^4+q^6\right)
\left(1+q^2+q^4+q^6+q^8\right)} \\
(u_1)_{23}&=&\frac{q^3 \sqrt{\left(1+\frac{1}{q^2}+q^2\right) \left(1+\frac{1}{q^4}+\frac{1}{q^2}+q^2+q^4\right)} \left(1-q^4+q^8\right)}{\left(1+q^2\right)
\left(1+q^4\right) \left(1+q^2+q^4+q^6+q^8\right)}\\
(u_1)_{32}&=& \frac{q^3 \sqrt{\left(1+\frac{1}{q^2}+q^2\right) \left(1+\frac{1}{q^4}+\frac{1}{q^2}+q^2+q^4\right)}
\left(1-q^4+q^8\right)}{\left(1+q^2\right) \left(1+q^4\right) \left(1+q^2+q^4+q^6+q^8\right)}\\
(u_1)_{34}&=& \frac{q^9
\sqrt{\left(1+\frac{1}{q^4}+\frac{1}{q^2}+q^2+q^4\right) \left(1+\frac{1}{q^6}+\frac{1}{q^4}+\frac{1}{q^2}+q^2+q^4+q^6\right)}}{\left(1+q^4\right)
\left(1+q^6\right) \left(1+q^2+q^4+q^6+q^8\right)}\\
(u_1)_{42}&=& -\frac{q^5 \left(1+q^2+q^4\right) \sqrt{\left(1+\frac{1}{q^2}+q^2\right) \left(1+\frac{1}{q^6}+\frac{1}{q^4}+\frac{1}{q^2}+q^2+q^4+q^6\right)}}{\left(1+q^2+q^4+q^6\right)
\left(1+q^2+q^4+q^6+q^8\right)}\\
(u_1)_{44}&=&-\frac{q^9}{\left(1+q^4\right) \left(1+q^6\right) \left(1+q^2+q^4+q^6+q^8\right)} \\
 (u_1)_{43}&=&\frac{q^9 \sqrt{\left(1+\frac{1}{q^4}+\frac{1}{q^2}+q^2+q^4\right) \left(1+\frac{1}{q^6}+\frac{1}{q^4}+\frac{1}{q^2}+q^2+q^4+q^6\right)}}{\left(1+q^4\right)
\left(1+q^6\right) \left(1+q^2+q^4+q^6+q^8\right)} 
\end{eqnarray*}

\begin{equation}
\mathcal{U}^{[7,2,0]}=\left(
\begin{array}{ccc}
 -\frac{q^4 \left(1+q^2+q^4\right)}{\left(1+q^4\right) \left(1+q^2+q^4+q^6+q^8\right)} & -\frac{q^2 \sqrt{\frac{1+q^2+q^4+q^6+q^8+q^{10}+q^{12}}{1+q^4+q^6+q^8+q^{12}}}}{1+q^4}
& (u_2)_{13}\\
 -\frac{q^2 \sqrt{\frac{1+q^2+q^4+q^6+q^8+q^{10}+q^{12}}{1+q^4+q^6+q^8+q^{12}}}}{1+q^4} & -1+\frac{q^4}{\left(1+q^4\right) \left(1-q^2+q^4\right)}
& (u_2)_{23} \\
 (u_2)_{31} &(u_2)_{32}& -\frac{q^6}{1+q^4+q^6+q^8+q^{12}} \\
\end{array}
\right)
\end{equation}
\begin{eqnarray*}
(u_2)_{13}&=&-\frac{\left(1+q^8\right) \left(1+q^2+q^4+q^6+q^8+q^{10}+q^{12}\right)}{\left(1+q^2+q^4+q^6+q^8\right) \sqrt{1+q^4+q^6+2 q^8+q^{10}+2 q^{12}+q^{14}+2
q^{16}+q^{18}+q^{20}+q^{24}}} \\
(u_2)_{23}&=& \frac{q^2 \sqrt{\frac{1+q^8}{1+q^2+q^4+q^6+q^8}}}{1-q^2+q^4}\\
(u_2)_{32}&=& \frac{q^2 \sqrt{\frac{1+q^8}{1+q^2+q^4+q^6+q^8}}}{1-q^2+q^4} \\
(u_2)_{31}&=&-\frac{\left(1+q^8\right) \left(1+q^2+q^4+q^6+q^8+q^{10}+q^{12}\right)}{\left(1+q^2+q^4+q^6+q^8\right) \sqrt{1+q^4+q^6+2 q^8+q^{10}+2 q^{12}+q^{14}+2
q^{16}+q^{18}+q^{20}+q^{24}}}
\end{eqnarray*}

\begin{equation}
X_3^{[6,3,0]}=\left(
\begin{array}{cccc}
 -\frac{1}{q^{12} \left(1+q^2+q^4+q^6\right)} & \frac{\sqrt{1+q^2+q^4}}{q^{11} \left(1+q^2+q^4+q^6\right)} &(x_3)_{13} &(x_3)_{14} \\
 -\frac{\sqrt{1+q^2+q^4}}{q^9 \left(1+q^2+q^4+q^6\right)} & \frac{1+2 q^2+2 q^4+q^6+2 q^8+2 q^{10}+q^{12}}{q^8 \left(1+q^2\right) \left(1+q^4\right)
\left(1+q^2+q^4+q^6+q^8\right)} &(x_3)_{23} & (x_3)_{23} \\
 -\frac{\sqrt{1+q^2+q^4+q^6+q^8}}{q^2+q^4+q^6+q^8} &(x_3)_{32} & \frac{q^4 \left(1+q^2-q^4+q^6+q^8\right)}{1+q^4+q^6+q^{10}} &(x_3)_{34} \\
 -\frac{q^9 \sqrt{1+q^2+q^4+q^6+q^8+q^{10}+q^{12}}}{1+q^2+q^4+q^6} & (x_3)_{42}& (x_3)_{43}
& (x_3)_{44}\\
\end{array}
\right)
\end{equation}
\begin{eqnarray*}
(x_3)_{13}&=& -\frac{\sqrt{1+q^2+q^4+q^6+q^8}}{q^8
\left(1+q^2+q^4+q^6\right)}\\
(x_3)_{14}&=& \frac{\sqrt{1+q^2+q^4+q^6+q^8+q^{10}+q^{12}}}{q^3+q^5+q^7+q^9}\\
(x_3)_{23}&=& \frac{\left(-1+q^4-q^8\right) \sqrt{1+q^2 \left(1+q^2\right) \left(2+q^2+2 q^4+q^6+q^8\right)}}{q^5 \left(1+q^2\right)
\left(1+q^4\right) \left(1+q^2+q^4+q^6+q^8\right)}\\
(x_3)_{24}&=&-\frac{q^2 \left(1+q^2+q^4\right) \sqrt{1+q^2 \left(1+q^2\right) \left(2+q^2+2 q^4+q^6+2 q^8+q^{10}+q^{12}\right)}}{\left(1+q^2\right)
\left(1+q^4\right) \left(1+q^2+q^4+q^6+q^8\right)}\\
(x_3)_{32}&=& \frac{\left(1-q^4+q^8\right) \sqrt{1+q^2 \left(1+q^2\right) \left(2+q^2+2 q^4+q^6+q^8\right)}}{q
\left(1+q^2\right) \left(1+q^4\right) \left(1+q^2+q^4+q^6+q^8\right)}\\
(x_3)_{34}&=&\frac{q^{13}
\sqrt{1+q^2 \left(1+q^2\right) \left(2+q^2+3 q^4+2 q^6+3 q^8+2 q^{10}+2 q^{12}+q^{14}+q^{16}\right)}}{\left(1+q^2\right) \left(1+q^4\right) \left(1-q^2+q^4\right)
\left(1+q^2+q^4+q^6+q^8\right)} \\
(x_3)_{42}&=& -\frac{q^{12} \left(1+q^2+q^4\right) \sqrt{1+q^2 \left(1+q^2\right) \left(2+q^2+2
q^4+q^6+2 q^8+q^{10}+q^{12}\right)}}{\left(1+q^2\right) \left(1+q^4\right) \left(1+q^2+q^4+q^6+q^8\right)}\\
(x_3)_{43}&=&-\frac{q^{19} \sqrt{1+q^2 \left(1+q^2\right)
\left(2+q^2+3 q^4+2 q^6+3 q^8+2 q^{10}+2 q^{12}+q^{14}+q^{16}\right)}}{\left(1+q^2\right) \left(1+q^4\right) \left(1-q^2+q^4\right) \left(1+q^2+q^4+q^6+q^8\right)}\\
(x_3)_{44}&=&-\frac{q^{30}}{\left(1+q^2\right) \left(1+q^4\right) \left(1-q^2+q^4\right) \left(1+q^2+q^4+q^6+q^8\right)} 
\end{eqnarray*}

\begin{equation*}
X_1^{[5,3,1]}=\left(
\begin{array}{ccc}
 \frac{1}{q^6+q^8+q^{10}} & -\frac{1}{q^5 \sqrt{1+q^2+q^4}} & \frac{\sqrt{1+q^2+q^4+q^6+q^8}}{q^2+q^4+q^6} \\
 \frac{1}{q^3 \sqrt{1+q^2+q^4}} & -\frac{1}{q^2}+\frac{1}{1+q^4} & -\frac{q^3 \sqrt{\frac{1+q^2+q^4+q^6+q^8}{1+q^2+q^4}}}{1+q^4} \\
 \frac{q^4 \sqrt{1+q^2+q^4+q^6+q^8}}{1+q^2+q^4} & \frac{q^7 \sqrt{\frac{1+q^2+q^4+q^6+q^8}{1+q^2+q^4}}}{1+q^4} & \frac{q^{14}}{1+q^2+2 q^4+q^6+q^8}
\\
\end{array}
\right)
\end{equation*}

\begin{equation}
X_2^{[7,2,0]}={\left(
\begin{array}{ccc}
 \frac{q^{22}}{1+q^4+q^6+q^8+q^{12}} & \frac{q^{12} \sqrt{\frac{1+q^8}{1+q^2+q^4+q^6+q^8}}}{1-q^2+q^4} & (x_2)_{13} \\
 -\frac{q^6 \sqrt{\frac{1+q^8}{1+q^2+q^4+q^6+q^8}}}{1-q^2+q^4} &\frac{-1+q^2-q^4+q^6-q^8}{q^2 \left(1+q^4\right) \left(1-q^2+q^4\right)} &(x_2)_{23}  \\
 (x_2)_{31} & \frac{\sqrt{\left(1+q^4+q^6+q^8+q^{12}\right)
\left(1+q^2+q^4+q^6+q^8+q^{10}+q^{12}\right)}}{q^8 \left(1+2 q^4+q^6+2 q^8+q^{10}+2 q^{12}+q^{16}\right)} & \frac{1+q^2+q^4}{q^{10} \left(1+q^4\right)
\left(1+q^2+q^4+q^6+q^8\right)} \\
\end{array}
\right)}
\end{equation}
where
\begin{eqnarray*}
(x_2)_{13}&=&-\frac{q^6 \sqrt{\frac{1+q^2+q^4+q^6+2 q^8+2
q^{10}+2 q^{12}+q^{14}+q^{16}+q^{18}+q^{20}}{1-q^2+q^4}}}{1+q^2+q^4+q^6+q^8}\\
(x_2)_{23}&=&-\frac{\sqrt{\left(1+q^4+q^6+q^8+q^{12}\right)
\left(1+q^2+q^4+q^6+q^8+q^{10}+q^{12}\right)}}{q^4 \left(1+2 q^4+q^6+2 q^8+q^{10}+2 q^{12}+q^{16}\right)}\\
(x_2)_{31}&=&-\frac{\sqrt{\frac{1+q^2+q^4+q^6+2 q^8+2 q^{10}+2 q^{12}+q^{14}+q^{16}+q^{18}+q^{20}}{1-q^2+q^4}}}{q^4+q^6+q^8+q^{10}+q^{12}}\\
\end{eqnarray*}

\section{Appendix B}{\label{app1}}

The weaving knot $W(3,n)$ whose $[r]=2, 3$ colored HOMFLY-PT worked out in chapter \ref{s4.int} (see in eqn. (\ref{HW2}) $\&$ (\ref{HW33})) can be compactly rewritten in the matrix form $(q^2,A^2)$:
As example [2]-colored  HOMFLY-PT of $W(3,2)$ knot  is
\begin{eqnarray*}
H_{[2]}^{W[3,2]}&=&\frac{1}{A^4 q^6}(-A^2+A^4+q^2-A^2 q^2-A^4 q^2+A^2 q^4-A^6 q^4+3 A^4 q^6-A^2 q^8+A^6 q^8\\&&
-A^4 q^{10}-A^6 q^{10}+A^8 q^{10}+A^4 q^{12}-A^6 q^{12}),
\end{eqnarray*}
and it can compactly rewritten in the matrix form $(q^2,A^2)$  
\begin{equation*}
H_{[2]}^{W[3,2]}=A^{-4} q^{-6} \left(
\begin{array}{ccccc}
  0 & -1 & 1 & 0 & 0 \\
 1 & -1 & -1 & 0 & 0 \\
 0 & 0 & 3 & 0 & 0 \\
 0 & -1 & 0 & 1 & 0 \\
 0 & 0 & -1 & -1 & 1 \\
 0 & 0 & 1 & -1 & 0 \\
\end{array}
\right).
\end{equation*}
Similarly, colored HOMFLY-PT for few other weaving knots listed in the matrix form $(q^2,A^2)$:
\begin{equation*}
H_{[2]}^{W[3,4]}=A^{-4} q^{-18} \left(
\begin{array}{ccccc}
 0 & -1 & 1 & 0 & 0 \\
 1 & 2 & -3 & 0 & 0 \\
 -3 & 3 & 1 & -1 & 0 \\
 0 & -10 & 7 & 2 & 0 \\
 9 & 2 & -14 & 3 & 0 \\
 -8 & 19 & 2 & -9 & 1 \\
 -7 & -20 & 27 & 3 & -3 \\
 15 & -15 & -26 & 16 & 0 \\
 -3 & 31 & -13 & -23 & 8 \\
 -9 & -8 & 47 & -8 & -9 \\
 8 & -23 & -13 & 31 & -3 \\
 0 & 16 & -26 & -15 & 15 \\
 -3 & 3 & 27 & -20 & -7 \\
 1 & -9 & 2 & 19 & -8 \\
 0 & 3 & -14 & 2 & 9 \\
 0 & 2 & 7 & -10 & 0 \\
 0 & -1 & 1 & 3 & -3 \\
 0 & 0 & -3 & 2 & 1 \\
 0 & 0 & 1 & -1 & 0 \\
\end{array}
\right),~H_{[2]}^{W[3,5]}=A^{-4} q^{-24} \left(
\begin{array}{ccccc}
 0 & -1 & 1 & 0 & 0 \\
 1 & 3 & -4 & 0 & 0 \\
 -4 & 2 & 3 & -1 & 0 \\
 2 & -15 & 10 & 3 & 0 \\
 12 & 13 & -26 & 2 & 0 \\
 -22 & 25 & 11 & -15 & 1 \\
 -2 & -58 & 46 & 12 & -4 \\
 45 & 5 & -76 & 24 & 2 \\
 -36 & 92 & 1 & -54 & 12 \\
 -29 & -87 & 129 & 8 & -21 \\
 66 & -58 & -116 & 85 & -1 \\
 -19 & 136 & -65 & -93 & 41 \\
 -40 & -40 & 189 & -40 & -40 \\
 41 & -93 & -65 & 136 & -19 \\
 -1 & 85 & -116 & -58 & 66 \\
 -21 & 8 & 129 & -87 & -29 \\
 12 & -54 & 1 & 92 & -36 \\
 2 & 24 & -76 & 5 & 45 \\
 -4 & 12 & 46 & -58 & -2 \\
 1 & -15 & 11 & 25 & -22 \\
 0 & 2 & -26 & 13 & 12 \\
 0 & 3 & 10 & -15 & 2 \\
 0 & -1 & 3 & 2 & -4 \\
 0 & 0 & -4 & 3 & 1 \\
 0 & 0 & 1 & -1 & 0 \\
\end{array}
\right)
\end{equation*}

{\tiny \begin{equation*}
H_{[2]}^{W[3,10]}=A^{-4} q^{-54} \left(\begin{array}{ccccc}
 0 & -1 & 1 & 0 & 0 \\
 1 & 8 & -9 & 0 & 0 \\
 -9 & -18 & 28 & -1 & 0 \\
 27 & -25 & -10 & 8 & 0 \\
 -3 & 182 & -161 & -18 & 0 \\
 -171 & -215 & 410 & -25 & 1 \\
 368 & -452 & -89 & 182 & -9 \\
 60 & 1550 & -1422 & -215 & 27 \\
 -1436 & -763 & 2654 & -452 & -3 \\
 2002 & -3631 & 249 & 1550 & -171 \\
 1201 & 6924 & -7730 & -763 & 368 \\
 -6748 & 283 & 10046 & -3630 & 60 \\
 5900 & -16340 & 4951 & 6925 & -1436 \\
 7228 & 18695 & -28254 & 274 & 2002 \\
 -20384 & 11318 & 24214 & -16348 & 1200 \\
 9951 & -47634 & 25873 & 18733 & -6749 \\
 24582 & 31349 & -73186 & 11346 & 5909 \\
 -42876 & 46700 & 36275 & -47744 & 7236 \\
 6553 & -97291 & 79874 & 31285 & -20421 \\
 56419 & 26407 & -138924 & 46951 & 9924 \\
 -64446 & 113504 & 23438 & -97179 & 24683 \\
 -13101 & -142395 & 171125 & 25948 & -42820 \\
 92427 & -16795 & -195296 & 113324 & 6340 \\
 -67075 & 189673 & -35612 & -141595 & 56325 \\
 -45008 & -143645 & 269458 & -16752 & -64053 \\
 108753 & -89895 & -196699 & 188829 & -13061 \\
 -44308 & 224367 & -128141 & -143909 & 91991 \\
 -67241 & -88626 & 313941 & -88626 & -67241 \\
 91991 & -143909 & -128141 & 224367 & -44308 \\
 -13061 & 188829 & -196699 & -89895 & 108753 \\
 -64053 & -16752 & 269458 & -143645 & -45008 \\
 56325 & -141595 & -35612 & 189673 & -67075 \\
 6340 & 113324 & -195296 & -16795 & 92427 \\
 -42820 & 25948 & 171125 & -142395 & -13101 \\
 24683 & -97179 & 23438 & 113504 & -64446 \\
 9924 & 46951 & -138924 & 26407 & 56419 \\
 -20421 & 31285 & 79874 & -97291 & 6553 \\
 7236 & -47744 & 36275 & 46700 & -42876 \\
 5909 & 11346 & -73186 & 31349 & 24582 \\
 -6749 & 18733 & 25873 & -47634 & 9951 \\
 1200 & -16348 & 24214 & 11318 & -20384 \\
 2002 & 274 & -28254 & 18695 & 7228 \\
 -1436 & 6925 & 4951 & -16340 & 5900 \\
 60 & -3630 & 10046 & 283 & -6748 \\
 368 & -763 & -7730 & 6924 & 1201 \\
 -171 & 1550 & 249 & -3631 & 2002 \\
 -3 & -452 & 2654 & -763 & -1436 \\
 27 & -215 & -1422 & 1550 & 60 \\
 -9 & 182 & -89 & -452 & 368 \\
 1 & -25 & 410 & -215 & -171 \\
 0 & -18 & -161 & 182 & -3 \\
 0 & 8 & -10 & -25 & 27 \\
 0 & -1 & 28 & -18 & -9 \\
 0 & 0 & -9 & 8 & 1 \\
 0 & 0 & 1 & -1 & 0 \\
\end{array}
\right),~
H_{[2]}^{W[3,11]}=A^{-4} q^{-60} \left(
\begin{array}{ccccc}
 0 & -1 & 1 & 0 & 0 \\
 1 & 9 & -10 & 0 & 0 \\
 -10 & -25 & 36 & -1 & 0 \\
 35 & -16 & -28 & 9 & 0 \\
 -20 & 234 & -189 & -25 & 0 \\
 -205 & -397 & 617 & -16 & 1 \\
 577 & -434 & -367 & 234 & -10 \\
 -158 & 2443 & -1923 & -397 & 35 \\
 -2060 & -2239 & 4753 & -434 & -20 \\
 3927 & -4832 & -1333 & 2443 & -205 \\
 486 & 13585 & -12408 & -2239 & 577 \\
 -11853 & -5003 & 21846 & -4832 & -158 \\
 14988 & -28037 & 1513 & 13584 & -2060 \\
 8635 & 46774 & -54332 & -5004 & 3927 \\
 -44032 & 5271 & 66368 & -28027 & 486 \\
 35635 & -102619 & 32053 & 46783 & -11852 \\
 43315 & 106204 & -169962 & 5224 & 14989 \\
 -114157 & 70990 & 137197 & -102655 & 8625 \\
 51703 & -260950 & 147528 & 106353 & -44041 \\
 130381 & 156476 & -393621 & 71083 & 35681 \\
 -216338 & 250236 & 182839 & -261319 & 43350 \\
 28407 & -484855 & 414452 & 156292 & -114296 \\
 274974 & 114379 & -689801 & 250986 & 51619 \\
 -303193 & 555289 & 101777 & -484577 & 130704 \\
 -65130 & -665659 & 830585 & 113126 & -216190 \\
 427192 & -92394 & -917444 & 554840 & 27806 \\
 -305233 & 878875 & -180468 & -663554 & 274736 \\
 -205261 & -652190 & 1251930 & -92336 & -302143 \\
 494008 & -412593 & -898236 & 876735 & -65072 \\
 -203589 & 1021418 & -591159 & -652754 & 426084 \\
 -305582 & -409517 & 1435653 & -409517 & -305582 \\
 426084 & -652754 & -591159 & 1021418 & -203589 \\
 -65072 & 876735 & -898236 & -412593 & 494008 \\
 -302143 & -92336 & 1251930 & -652190 & -205261 \\
 274736 & -663554 & -180468 & 878875 & -305233 \\
 27806 & 554840 & -917444 & -92394 & 427192 \\
 -216190 & 113126 & 830585 & -665659 & -65130 \\
 130704 & -484577 & 101777 & 555289 & -303193 \\
 51619 & 250986 & -689801 & 114379 & 274974 \\
 -114296 & 156292 & 414452 & -484855 & 28407 \\
 43350 & -261319 & 182839 & 250236 & -216338 \\
 35681 & 71083 & -393621 & 156476 & 130381 \\
 -44041 & 106353 & 147528 & -260950 & 51703 \\
 8625 & -102655 & 137197 & 70990 & -114157 \\
 14989 & 5224 & -169962 & 106204 & 43315 \\
 -11852 & 46783 & 32053 & -102619 & 35635 \\
 486 & -28027 & 66368 & 5271 & -44032 \\
 3927 & -5004 & -54332 & 46774 & 8635 \\
 -2060 & 13584 & 1513 & -28037 & 14988 \\
 -158 & -4832 & 21846 & -5003 & -11853 \\
 577 & -2239 & -12408 & 13585 & 486 \\
 -205 & 2443 & -1333 & -4832 & 3927 \\
 -20 & -434 & 4753 & -2239 & -2060 \\
 35 & -397 & -1923 & 2443 & -158 \\
 -10 & 234 & -367 & -434 & 577 \\
 1 & -16 & 617 & -397 & -205 \\
 0 & -25 & -189 & 234 & -20 \\
 0 & 9 & -28 & -16 & 35 \\
 0 & -1 & 36 & -25 & -10 \\
 0 & 0 & -10 & 9 & 1 \\
 0 & 0 & 1 & -1 & 0 \\
\end{array}
\right)
\end{equation*}}

\begin{equation*}
H_{[3]}^{W[3,2]}=A^{-6} q^{-14} \left(
\begin{array}{ccccccc}
 0 & -1 & 1 & 0 & 0 & 0 & 0 \\
 1 & -1 & 1 & -1 & 0 & 0 & 0 \\
 0 & -1 & 0 & 1 & 0 & 0 & 0 \\
 0 & 1 & -1 & 0 & 0 & 0 & 0 \\
 0 & 0 & 0 & -2 & 1 & 0 & 0 \\
 0 & 0 & 3 & -3 & 0 & 0 & 0 \\
 0 & -1 & 2 & 0 & -1 & 0 & 0 \\
 0 & 0 & 0 & 3 & 0 & 0 & 0 \\
 0 & 0 & -1 & 0 & 2 & -1 & 0 \\
 0 & 0 & 0 & -3 & 3 & 0 & 0 \\
 0 & 0 & 1 & -2 & 0 & 0 & 0 \\
 0 & 0 & 0 & 0 & -1 & 1 & 0 \\
 0 & 0 & 0 & 1 & 0 & -1 & 0 \\
 0 & 0 & 0 & -1 & 1 & -1 & 1 \\
 0 & 0 & 0 & 0 & 1 & -1 & 0 \\
\end{array}
\right)
\end{equation*}

{\tiny \begin{equation*}
H_{[3]}^{W[3,4]}=A^{-6} q^{-38} \left(
\begin{array}{ccccccc}
 0 & -1 & 1 & 0 & 0 & 0 & 0 \\
 1 & 2 & -2 & -1 & 0 & 0 & 0 \\
 -3 & 2 & -2 & 3 & 0 & 0 & 0 \\
 0 & -3 & 4 & -1 & 0 & 0 & 0 \\
 6 & -11 & 8 & -4 & 1 & 0 & 0 \\
 4 & 4 & -6 & 0 & -2 & 0 & 0 \\
 -11 & 27 & -24 & 10 & -2 & 0 & 0 \\
 -17 & 6 & 8 & 0 & 3 & 0 & 0 \\
 17 & -44 & 48 & -27 & 7 & -1 & 0 \\
 28 & -39 & 11 & 2 & -4 & 2 & 0 \\
 -7 & 55 & -82 & 49 & -18 & 2 & 0 \\
 -42 & 79 & -59 & 15 & 9 & -2 & 0 \\
 -6 & -27 & 98 & -92 & 36 & -9 & 0 \\
 42 & -114 & 126 & -53 & 0 & 3 & 1 \\
 17 & -14 & -73 & 124 & -70 & 19 & -3 \\
 -31 & 113 & -190 & 136 & -26 & -2 & 0 \\
 -21 & 59 & -10 & -114 & 107 & -36 & 5 \\
 16 & -79 & 200 & -201 & 75 & -14 & 3 \\
 20 & -69 & 71 & 45 & -122 & 64 & -9 \\
 -11 & 37 & -151 & 263 & -151 & 37 & -11 \\
 -9 & 64 & -122 & 45 & 71 & -69 & 20 \\
 3 & -14 & 75 & -201 & 200 & -79 & 16 \\
 5 & -36 & 107 & -114 & -10 & 59 & -21 \\
 0 & -2 & -26 & 136 & -190 & 113 & -31 \\
 -3 & 19 & -70 & 124 & -73 & -14 & 17 \\
 1 & 3 & 0 & -53 & 126 & -114 & 42 \\
 0 & -9 & 36 & -92 & 98 & -27 & -6 \\
 0 & -2 & 9 & 15 & -59 & 79 & -42 \\
 0 & 2 & -18 & 49 & -82 & 55 & -7 \\
 0 & 2 & -4 & 2 & 11 & -39 & 28 \\
 0 & -1 & 7 & -27 & 48 & -44 & 17 \\
 0 & 0 & 3 & 0 & 8 & 6 & -17 \\
 0 & 0 & -2 & 10 & -24 & 27 & -11 \\
 0 & 0 & -2 & 0 & -6 & 4 & 4 \\
 0 & 0 & 1 & -4 & 8 & -11 & 6 \\
 0 & 0 & 0 & -1 & 4 & -3 & 0 \\
 0 & 0 & 0 & 3 & -2 & 2 & -3 \\
 0 & 0 & 0 & -1 & -2 & 2 & 1 \\
 0 & 0 & 0 & 0 & 1 & -1 & 0 \\
\end{array}
\right),~ 
H_{[3]}^{W[3,5]}=A^{-6} q^{-50} \left(
\begin{array}{ccccccc}
 0 & 1 & -1 & 0 & 0 & 0 & 0 \\
 -1 & -3 & 3 & 1 & 0 & 0 & 0 \\
 4 & -1 & 1 & -4 & 0 & 0 & 0 \\
 -2 & 6 & -7 & 3 & 0 & 0 & 0 \\
 -8 & 9 & -6 & 6 & -1 & 0 & 0 \\
 2 & -18 & 17 & -4 & 3 & 0 & 0 \\
 21 & -27 & 23 & -18 & 1 & 0 & 0 \\
 4 & 40 & -44 & 7 & -7 & 0 & 0 \\
 -58 & 73 & -57 & 48 & -7 & 1 & 0 \\
 -14 & -49 & 71 & -21 & 16 & -3 & 0 \\
 98 & -184 & 146 & -85 & 26 & -1 & 0 \\
 72 & 27 & -94 & 25 & -36 & 6 & 0 \\
 -145 & 329 & -316 & 179 & -57 & 10 & 0 \\
 -167 & 118 & 52 & -35 & 50 & -16 & -1 \\
 150 & -481 & 563 & -333 & 126 & -29 & 4 \\
 287 & -386 & 157 & -19 & -66 & 29 & -2 \\
 -91 & 482 & -793 & 589 & -249 & 64 & -8 \\
 -366 & 724 & -571 & 190 & 52 & -30 & 1 \\
 -22 & -308 & 872 & -868 & 448 & -142 & 20 \\
 386 & -962 & 1111 & -619 & 65 & 27 & 7 \\
 116 & -53 & -598 & 1036 & -699 & 248 & -50 \\
 -304 & 987 & -1532 & 1169 & -352 & 45 & -13 \\
 -180 & 365 & 72 & -863 & 915 & -412 & 79 \\
 207 & -763 & 1638 & -1749 & 812 & -197 & 52 \\
 174 & -564 & 508 & 334 & -863 & 541 & -130 \\
 -114 & 461 & -1314 & 1963 & -1314 & 461 & -114 \\
 -130 & 541 & -863 & 334 & 508 & -564 & 174 \\
 52 & -197 & 812 & -1749 & 1638 & -763 & 207 \\
 79 & -412 & 915 & -863 & 72 & 365 & -180 \\
 -13 & 45 & -352 & 1169 & -1532 & 987 & -304 \\
 -50 & 248 & -699 & 1036 & -598 & -53 & 116 \\
 7 & 27 & 65 & -619 & 1111 & -962 & 386 \\
 20 & -142 & 448 & -868 & 872 & -308 & -22 \\
 1 & -30 & 52 & 190 & -571 & 724 & -366 \\
 -8 & 64 & -249 & 589 & -793 & 482 & -91 \\
 -2 & 29 & -66 & -19 & 157 & -386 & 287 \\
 4 & -29 & 126 & -333 & 563 & -481 & 150 \\
 -1 & -16 & 50 & -35 & 52 & 118 & -167 \\
 0 & 10 & -57 & 179 & -316 & 329 & -145 \\
 0 & 6 & -36 & 25 & -94 & 27 & 72 \\
 0 & -1 & 26 & -85 & 146 & -184 & 98 \\
 0 & -3 & 16 & -21 & 71 & -49 & -14 \\
 0 & 1 & -7 & 48 & -57 & 73 & -58 \\
 0 & 0 & -7 & 7 & -44 & 40 & 4 \\
 0 & 0 & 1 & -18 & 23 & -27 & 21 \\
 0 & 0 & 3 & -4 & 17 & -18 & 2 \\
 0 & 0 & -1 & 6 & -6 & 9 & -8 \\
 0 & 0 & 0 & 3 & -7 & 6 & -2 \\
 0 & 0 & 0 & -4 & 1 & -1 & 4 \\
 0 & 0 & 0 & 1 & 3 & -3 & -1 \\
 0 & 0 & 0 & 0 & -1 & 1 & 0 \\
\end{array}
\right)
\end{equation*}}

\section{Appendix C}{\label{app2}}

\begin{equation}
\hat{\bf N}^{W[3,2]}_{[ 2]}=\left(
\begin{array}{c|ccccccc}
\hline
 k/m&-6 & -4 & -2 & 0 & 2 & 4 & 6 \\
\hline
 0&1 & -3 & 4 & -6 & 9 & -7 & 2 \\
1& 0 & -1 & 2 & -5 & 9 & -6 & 1 \\
2& 0 & 0 & 0 & -1 & 2 & -1 & 0 \\
\end{array}
\right)
\end{equation}

\begin{equation}
\hat{\bf N}^{W[3,5]}_{[ 2]}=\left(
\begin{array}{c|ccccccc}
\hline
 k/m&-6& -4 & -2 & 0 & 2 & 4 & 6 \\
\hline 
0&5 & -15 & 20 & -30 & 45 & -35 & 10 \\
1& -20 & 85 & -170 & 195 & -135 & 60 & -15 \\
2& -106 & 441 & -840 & 1125 & -1130 & 676 & -166 \\
3& -115 & 485 & -865 & 1400 & -1860 & 1245 & -290 \\
4& 58 & -304 & 893 & -895 & -17 & 319 & -54 \\
5& 220 & -1112 & 2721 & -3910 & 3407 & -1728 & 402 \\
6& 190 & -1053 & 2594 & -4360 & 4748 & -2667 & 548 \\
7& 75 & -504 & 1287 & -2580 & 3268 & -1874 & 328 \\
8& 14 & -132 & 354 & -901 & 1300 & -737 & 102 \\
9& 1 & -18 & 51 & -186 & 302 & -166 & 16 \\
10& 0 & -1 & 3 & -21 & 38 & -20 & 1 \\
11& 0 & 0 & 0 & -1 & 2 & -1 & 0 \\
\end{array}
\right)
\end{equation}

\begin{equation}
\hat{\bf N}^{W[3,10]}_{[ 2]}=\left(
\begin{array}{c|ccccccc}
\hline
 k/m&-6& -4 & -2 & 0 & 2 & 4 & 6 \\
\hline
0&15 & -45 & 60 & -90 & 135 & -105 & 30 \\
1& -60 & 705 & -2310 & 3285 & -2205 & 630 & -45 \\
2& -238 & 318 & 30 & 725 & -2340 & 2223 & -718 \\
3& -230 & -9535 & 40480 & -59150 & 35020 & -5455 & -1130 \\
4& -5276 & 18038 & -10726 & -8750 & 1419 & 7782 & -2487 \\
5& -14765 & 146944 & -442897 & 622355 & -446589 & 151571 & -16619 \\
6& 79740 & -198204 & -1608 & 289370 & -123746 & -101606 & 56054 \\
7& 508927 & -2324454 & 4616391 & -5906820 & 5390964 & -2958324 & 673316 \\
8& 1028581 & -4926087 & 10581436 & -15418579 & 15626039 & -8863136 & 1971746 \\
9& 391793 & -1556984 & 2518630 & -6792357 & 12260745 & -8784777 & 1962950 \\
10& -2012429 & 10695435 & -26645018 & 34658435 & -24266624 & 9735414 & -2165213 \\
11& -3782274 & 19693607 & -47993117 & 72809354 & -69531190 & 36832395 & -8028775 \\
12& -1512571 & 8341895 & -20569240 & 43380863 & -58198567 & 35897456 & -7339836 \\
13& 4189988 & -21105464 & 50727994 & -60798221 & 36574589 & -13279099 & 3690213 \\
14& 8727473 & -46175793 & 112341954 & -166703357 & 154882064 & -80507746 & 17435405 \\
15& 8982577 & -49903892 & 122763600 & -201133018 & 212030754 & -116092856 & 23352835 \\
16& 6117458 & -36148228 & 90092284 & -161044360 & 185210485 & -103620634 & 19392995 \\
17& 2978166 & -19030475 & 48156926 & -94240005 & 116596085 & -65816480 & 11355783 \\
18& 1065351 & -7510739 & 19336164 & -41833861 & 55278128 & -31237740 & 4902697 \\
19& 281585 & -2245022 & 5890311 & -14299136 & 20085108 & -11298452 & 1585606 \\
20& 54437 & -506753 & 1356948 & -3772701 & 5613530 & -3129606 & 384145 \\
21& 7490 & -85090 & 232800 & -762716 & 1198736 & -660014 & 68794 \\
22& 695 & -10309 & 28842 & -116056 & 192190 & -104205 & 8843 \\
23& 39 & -852 & 2439 & -12866 & 22400 & -11932 & 772 \\
24& 1 & -43 & 126 & -981 & 1792 & -936 & 41 \\
25& 0 & -1 & 3 & -46 & 88 & -45 & 1 \\
26& 0 & 0 & 0 & -1 & 2 & -1 & 0 \\
\end{array}
\right)
\end{equation}

\begin{equation}
\hat{\bf N}^{W[3,2]}_{[1,1]}=\left(
\begin{array}{c|ccccccc}
\hline
 k/m&-6& -4 & -2 & 0 & 2 & 4 & 6 \\
\hline
 0&-2 & 7 & -9 & 6 & -4 & 3 & -1 \\
 1&-1 & 6 & -9 & 5 & -2 & 1 & 0 \\
 2&0 & 1 & -2 & 1 & 0 & 0 & 0 \\
\end{array}
\right)
\end{equation}

\begin{equation}
\hat{\bf N}^{W[3,5]}_{[1,1]}=\left(
\begin{array}{c|ccccccc}
\hline
 k/m&-6& -4 & -2 & 0 & 2 & 4 & 6 \\
\hline
0& -10 & 35 & -45 & 30 & -20 & 15 & -5 \\
1& 15 & -60 & 135 & -195 & 170 & -85 & 20 \\
2& 166 & -676 & 1130 & -1125 & 840 & -441 & 106 \\
3& 290 & -1245 & 1860 & -1400 & 865 & -485 & 115 \\
4& 54 & -319 & 17 & 895 & -893 & 304 & -58 \\
5& -402 & 1728 & -3407 & 3910 & -2721 & 1112 & -220 \\
6& -548 & 2667 & -4748 & 4360 & -2594 & 1053 & -190 \\
7& -328 & 1874 & -3268 & 2580 & -1287 & 504 & -75 \\
8& -102 & 737 & -1300 & 901 & -354 & 132 & -14 \\
9& -16 & 166 & -302 & 186 & -51 & 18 & -1 \\
10& -1 & 20 & -38 & 21 & -3 & 1 & 0 \\
11& 0 & 1 & -2 & 1 & 0 & 0 & 0 \\
\end{array}
\right)
\end{equation}

\begin{equation}
\hat{\bf N}^{W[3,10]}_{[1,1]}=\left(
\begin{array}{c|ccccccc}
\hline
 k/m&-6& -4 & -2 & 0 & 2 & 4 & 6 \\
\hline
 0&-30 & 105 & -135 & 90 & -60 & 45 & -15 \\
 1&45 & -630 & 2205 & -3285 & 2310 & -705 & 60 \\
 2&718 & -2223 & 2340 & -725 & -30 & -318 & 238 \\
 3&1130 & 5455 & -35020 & 59150 & -40480 & 9535 & 230 \\
 4&2487 & -7782 & -1419 & 8750 & 10726 & -18038 & 5276 \\
 5&16619 & -151571 & 446589 & -622355 & 442897 & -146944 & 14765 \\
 6&-56054 & 101606 & 123746 & -289370 & 1608 & 198204 & -79740 \\
 7&-673316 & 2958324 & -5390964 & 5906820 & -4616391 & 2324454 & -508927 \\
 8&-1971746 & 8863136 & -15626039 & 15418579 & -10581436 & 4926087 & -1028581 \\
 9&-1962950 & 8784777 & -12260745 & 6792357 & -2518630 & 1556984 & -391793 \\
 10&2165213 & -9735414 & 24266624 & -34658435 & 26645018 & -10695435 & 2012429 \\
 11&8028775 & -36832395 & 69531190 & -72809354 & 47993117 & -19693607 & 3782274 \\
 12&7339836 & -35897456 & 58198567 & -43380863 & 20569240 & -8341895 & 1512571 \\
 13&-3690213 & 13279099 & -36574589 & 60798221 & -50727994 & 21105464 & -4189988 \\
 14&-17435405 & 80507746 & -154882064 & 166703357 & -112341954 & 46175793 & -8727473 \\
 15&-23352835 & 116092856 & -212030754 & 201133018 & -122763600 & 49903892 & -8982577 \\
 16&-19392995 & 103620634 & -185210485 & 161044360 & -90092284 & 36148228 & -6117458 \\
 17&-11355783 & 65816480 & -116596085 & 94240005 & -48156926 & 19030475 & -2978166 \\
 18&-4902697 & 31237740 & -55278128 & 41833861 & -19336164 & 7510739 & -1065351 \\
 19&-1585606 & 11298452 & -20085108 & 14299136 & -5890311 & 2245022 & -281585 \\
 20&-384145 & 3129606 & -5613530 & 3772701 & -1356948 & 506753 & -54437 \\
 21&-68794 & 660014 & -1198736 & 762716 & -232800 & 85090 & -7490 \\
 22&-8843 & 104205 & -192190 & 116056 & -28842 & 10309 & -695 \\
 23&-772 & 11932 & -22400 & 12866 & -2439 & 852 & -39 \\
 24&-41 & 936 & -1792 & 981 & -126 & 43 & -1 \\
 25&-1 & 45 & -88 & 46 & -3 & 1 & 0 \\
 26&0 & 1 & -2 & 1 & 0 & 0 & 0 \\
\end{array}
\right)
\end{equation}

\end{document}